%% file: wrapper.tex
\documentclass[%
 reprint,
 amsmath,amssymb,
 aps,
 twocolumn,
 superscriptaddress, 
 citeautoscript,
 compress,
]{revtex4-2}

\usepackage{graphicx}
\usepackage{dcolumn}
\usepackage{bm}
\usepackage{xcolor}
\usepackage{cancel}

\newcommand{\bg}{\boldsymbol{g}}

\newcommand{\av}[1]{\langle #1 \rangle }
\usepackage{soul}
\usepackage{etoolbox}

\DeclareMathOperator{\Tr}{Tr}

\usepackage{float}
\usepackage{ushort, mathtools}
\usepackage[
colorlinks=true, 
linkcolor=blue, 
citecolor=blue, 
urlcolor=blue, 
]{hyperref}

\begin{document}

\input{Main.tex}

\clearpage
\newpage
\onecolumngrid
\input{SI.tex}

\end{document}

%% file: Main.tex
\title{A High-Order Cumulant Extension of Quasi-Linkage Equilibrium}

\author{Kai S. Shimagaki}%
\thanks{Corresponding author}
 \email{kais@pitt.edu} %
 \affiliation{Department of Physics and Astronomy, and}
 \affiliation{Department of Computational and Systems Biology, University of Pittsburgh School of Medicine, USA.}

\author{Jorge Fernandez-de-Cossio-Diaz}%
\affiliation{Institut de Physique Théorique, Université Paris-Saclay, CNRS, CEA, Gif-sur-Yvette, France.}
\author{Mauro Pastore}
\affiliation{The Abdus Salam International Centre for Theoretical Physics (ICTP), Strada Costiera 11, 34151 Trieste, Italy}
\author{R\'emi Monasson}
\affiliation{Laboratoire de Physique de \'Ecole Normale Sup\'erieure, France}
\affiliation{PSL \& CNRS UMR 8023, France}
\affiliation{Sorbonne Universit\'e, 7005 Paris, France}
\author{Simona Cocco}
\affiliation{Laboratoire de Physique de \'Ecole Normale Sup\'erieure, France}
\affiliation{PSL \& CNRS UMR 8023, France}
\affiliation{Sorbonne Universit\'e, 7005 Paris, France}
\author{John P. Barton}%
\thanks{Corresponding author}
\email{jpbarton@pitt.edu} 
\affiliation{Department of Physics and Astronomy, and}
 \affiliation{Department of Computational and Systems Biology, University of Pittsburgh School of Medicine, USA.}
 
%



\begin{abstract}
A central question in evolutionary biology is how to quantitatively understand the dynamics of genetically diverse populations.
Modeling the genotype distribution is challenging, as it ultimately requires tracking all correlations (or cumulants) among alleles at different loci. The quasi-linkage equilibrium (QLE) approximation simplifies this by assuming that correlations between alleles at different loci are weak – i.e., low linkage disequilibrium – allowing their dynamics to be modeled perturbatively.
However, QLE breaks down under strong selection, significant epistatic interactions, or weak recombination.
We extend the multilocus QLE framework to allow cumulants up to order $K$ to evolve dynamically, while higher-order cumulants ($>K$) are assumed to equilibrate rapidly. This extended QLE (exQLE) framework yields a general equation of motion for cumulants up to order $K$, which parallels the standard QLE dynamics (recovered when $K = 1$). In this formulation, cumulant dynamics are driven by the gradient of average fitness, mediated by a geometrically interpretable matrix that stems from competition among genotypes.
Our analysis shows that the exQLE with $K=2$ accurately captures cumulant dynamics even when the fitness function includes higher-order (e.g., third- or fourth-order) epistatic interactions, capabilities that standard QLE lacks. We also applied the exQLE framework to infer fitness parameters from temporal sequence data. Overall, exQLE provides a systematic and interpretable approximation scheme, leveraging analytical cumulant dynamics and reducing complexity by progressively truncating higher-order cumulants.
\end{abstract}

\maketitle


\section*{Introduction}\label{sec:introduction}
One of the central questions in evolutionary biology is to understand how populations evolve under natural selection and other evolutionary forces \cite{wright1931evolution, fisher1999genetical, walsh2018evolution, crow2017introduction}. 
The aim is to model the dynamics of genotype distribution over time.
A major challenge arises from the non-random association of alleles (e.g., nucleotides) at different loci (positions in a linear sequence), known as linkage disequilibrium (LD) \cite{lewontin1960evolutionary}.
In the absence of LD, a state known as linkage equilibrium, each allele evolves independently and its dynamics can be solved exactly \cite{wright1931evolution}. However, this condition is overly restrictive. Indeed, natural selection can induce LD, which causes the dynamics of different alleles to become correlated.
A general description of population evolution requires accounting for an exponentially increasing number of allele combinations across loci, which are coupled through LD \cite{barton1989evolutionary, crow2017introduction}.

Generally, the time scale of change in LD is faster than the change in allele frequency. As a result, LD rapidly reaches equilibrium and is determined by allele frequency. This state, where LD values are small and stable while allele frequencies evolve, is known as quasi-linkage equilibrium (QLE) \cite{crow2017introduction, kimura1965attainment}.
In contrast to linkage equilibrium, QLE can naturally occur under conditions of weak selection and/or weak epistasis, coupled with high recombination rates. This results in distinct time scales: the slow dynamics of individual allele frequency, governed by selection (with a time scale of $1/s$, where $s$ is the selection strength), and the rapid decay of linkage disequilibrium (with a time scale of $1/r$, where $r$ is the recombination rate). The slow dynamics are driven by gradients of average fitness with respect to cumulants \cite{kimura1965attainment, barton1995general, neher2009competition, neher2011statistical}. 
Practically, QLE serves as a useful platform to investigate the collective evolution of alleles in multilocus systems, as it simplifies the mathematical structure, and reduces the dimensionality of genotype distribution \cite{neher2011statistical}. 

QLE has primarily been examined within a two-locus, two-allele framework, initially proposed by Kimura \cite{kimura1965attainment} and further investigated by Nagylaki, Barton and colleagues \cite{nagylaki1974quasilinkage, barton2005evolution}.  
Barton and Turelli developed a framework for describing evolution in centered moments \cite{barton1991natural,turelli1994genetic}, encompassing second and higher orders under arbitrary selection and recombination, which was later generalized by Kirkpatrick et al. \cite{kirkpatrick2002general}. Nagylaki et al. also focused on the evolution of multilocus systems, rigorously examining the sufficient conditions for convergence to equilibria or QLE manifolds using a small-epistasis perturbation theory \cite{nagylaki1993evolution, nagylaki1999convergence}.
Recently, Neher and Shraiman further developed the QLE theory for multilocus systems, building on the foundational work by Barton and Turelli \cite{barton1991natural, turelli1994genetic}. This has made the QLE theory more conceptually and analytically streamlined \cite{neher2011statistical}, elucidating that the cumulant dynamics are driven by the gradient of average fitenss. Recent work leveraged the analytically tractable QLE's cumulant dynamics to infer fitness parameters, an important challenge in evolutionary biology \cite{zeng2020inferring, zeng2020global}. 

Strong selection can at least transiently break down QLE. During a selective sweep, linked loci can exhibit positive LD as the selected loci drag other linked loci, known as genetic hitchhiking \cite{smith1974hitch} (see ref.~\cite{barton2000genetic} for a broader discussion). In contrast, negative LD may arise between selected loci, known as the Hill-Robertson interference effect \cite{hill1966effect} (see also ref.~\cite{felsenstein1974evolutionary}). These dynamics violate QLE, but equilibrium may re-establish after the sweep concludes.
However, epistatic interactions between different loci can permanently violate the QLE assumptions \cite{barton1995general, nagylaki1999convergence}. 
For example, if pairs of loci are selected due to epistatic interactions, then the pairwise LD or pairwise cumulants of alleles at different loci cannot relax to equilibria rapidly. As a result, the decay times of the pairwise cumulants become comparable to the timescales of individual allele frequencies.

Theoretical and computational studies offer insights into how epistasis arises \cite{husain2020physical, kryazhimskiy2021emergence}, the structure of epistasis \cite{poelwijk2016context, otwinowski2018inferring, zhou2022higher, rijal2025inferring}, and how epistasis influences evolutionary dynamics \cite{lyons2020idiosyncratic, reddy2021global} (See also recent reviews \cite{johnson2023epistasis, diaz2023global}).
However, it remains challenging to understand how populations evolve in the presence of higher-order epistasis.
Recent studies have revealed the prevalence and complex patterns of epistasis \cite{chou2011diminishing, khan2011negative, kryazhimskiy2014global, olson2014comprehensive, starr2018pervasive, park2022epistatic, bakerlee2022idiosyncratic}, including the presence of hierarchical higher-order epistasis in various contexts \cite{sailer2017high, poelwijk2019learning, phillips2021binding, moulana2023landscape, phillips2023hierarchical, yang2019higher, buda2023pervasive, schulz2025epistatic}. 
This underscores the need to extend QLE theory to accommodate a prevalent and broader range of epistasis.

Here, we propose an extended QLE (exQLE) framework that relaxes the assumptions that while individual allele frequencies evolve slowly, second and higher-order cumulants are small and converge rapidly to equilibria. Instead, we allow cumulants up to an arbitrary order $K$ to evolve dynamically while assuming those of order $K+1$ and above remain small and rapidly reach equilibrium. We first derived a general expression for cumulant dynamics under an arbitrary order of cumulants and genotype distributions, showing that their evolution is driven by the gradient of the average fitness function with respect to the cumulants. This forms the basis of the exQLE formulation, allowing for systematic relaxation of the assumption that cumulants above the $K=1$ order are effectively in steady state. 

As an example, we analyze the $K=2$ case, which is the simplest extension of QLE, demonstrating the dynamics of first- and second-order cumulants.  
These expressions describe how fitness parameters influence the cumulant dynamics and can be used to infer these fitness parameters. 
Interestingly, the expressions for the cumulant dynamics from exQLE match exactly with those derived by projecting genotype dynamics onto the space of cumulants in the diffusion approximation of the Wright-Fisher (WF) process \cite{sohail2021mpl,sohail2022inferring}.
Furthermore, the exQLE framework, whose cumulant dynamics are fully characterized by combinations of cumulants, readily provides a systematic approach to approximate the dynamics by progressively reducing cumulants in an order-by-order manner. 
This systematic approach reproduces the previously reported QLE-based epistasis inference method with a Gaussian closure (GC) scheme \cite{mauri2021gaussian}, along with alternative inference methods.

\section*{Evolution of genetic traits and Quasi-Linkage Equilibrium theory}
\phantomsection
\label{sec:problem_setup}
Here, we consider the evolution of a population of individuals, described by a probability distribution of genotypes, $P(\boldsymbol{g}, t)$ with a binary genotype $\boldsymbol{g}\in\{-1, +1 \}^L$ of length $L$. 
The average of an arbitrary genetic trait across a temporal genetic distribution is defined as $\langle G \rangle := \sum_{\boldsymbol{g}} P(\boldsymbol{g}, t) G(\boldsymbol{g})$. 
For the sake of simplicity, the time dependency ($t$) has been omitted. 
The principle governing population evolution is that fitter genotypes produce more offspring, leading to their increased frequency in the next generation. 
Let $F(\boldsymbol{g})$ denotes a fitness function, mapping genotypes to Malthusian fitness \cite{murray2007mathematical, ewens2004mathematical}, then the genotype dynamics can then be expressed as $P(g,t+\Delta t) = \frac{e^{\Delta t F(g)}}{\langle e^{\Delta t F(g)} \rangle} P(g, t)$ over a time period $\Delta t$. 
When the selection is small such that $|\log\left(e^{\Delta t F} / \langle e^{\Delta t F} \rangle \right)| \ll 1$, the expected genotype follows $P(g,t+\Delta t) \simeq P(g,t) + \Delta t [F(g) - \langle F \rangle]P(g,t)$. 
Considering mutation and recombination effects, which operate on individuals and introduce genetic variation, the dynamics of the genotype distribution is described by the following master equation:
\begin{align}
\begin{aligned}\label{eq:master_eq}
    &\dot{P}(\bg,t) = 
    \left[F(\bg)- \av{F}\right] P(\bg, t)\\
    &\quad +\mu \sum_{\boldsymbol{g}';\, d(\boldsymbol{g}, \boldsymbol{g}')=1} \left[P(\bg' , t) - P(\bg,t)\right]  \\
    &\quad + r \sum_{\boldsymbol{g}', \boldsymbol{g}''} R(\boldsymbol{g}\mid\boldsymbol{g}', \boldsymbol{g}'') P(\boldsymbol{g}',t )P(\boldsymbol{g}'',t) - rP(\boldsymbol{g},t)
\end{aligned}
\end{align}
where $\mu$, $r$ are, respectively, mutation and recombination rates, $d(\boldsymbol{g}, \boldsymbol{g}')$ represents the Hamming distance, and 
$R(\boldsymbol{g}\mid\boldsymbol{g}', \boldsymbol{g}'')$ is the probability that genotypes $\boldsymbol{g}'$ and $\boldsymbol{g}''$ produce genotype $\boldsymbol{g}$ through recombination. 

For simplicity, we temporarily ignore the contributions of mutation and recombination ($\mu=0$ and $r=0$), as their effects do not alter the following discussion and are explicitly detailed in refs.~\cite{neher2011statistical, sohail2022inferring}. 
Given this dynamical rule, the equation of motion for the average arbitrary trait or arbitrary function of $\boldsymbol{g}$ denoted as $G$, is given as:$\frac{d}{dt}\langle G \rangle  = \mathrm{Cov}(F,G) $\,, 
which is known as Price's equation \cite{price1972fisher} or ``second theorem'' of natural selection\cite{robertson1966mathematical}, and can be viewed as a generalization of Fisher's ``fundamental theorem'' \cite{fisher1999genetical, price1972fisher}.
Throughout this paper, we assume the fitness function is expressed as 
\begin{equation}\label{eq:fitness}
    F(\boldsymbol{g}) =  \sum_i s_i g_i + \sum_{i<j} s_{ij} g_i g_j + \ldots\,,
\end{equation}
where $s_i$ and $s_{ij}$ are time-independent coefficients characterizing the effects of a single mutation at site $i$ (selection coefficient) and a double mutation at site $i$ and $j$ (pairwise epistatic coefficient). 
To describe dynamics of cumulants below, we define the following cumulant generating function under an arbitrary genotype distribution, $P(\boldsymbol{g},t)$, parameterized by $\boldsymbol{\phi}=(\phi_1,\ldots,\phi_L)^\top$\,, given as: 
\begin{equation}\label{eq:cumlant_generating}
    \Phi(\boldsymbol{\phi},t) := \log\left(\sum_{\boldsymbol{g}} P(\boldsymbol{g},t)\ e^{\boldsymbol{\phi}^\top \boldsymbol{g}}\right)\,.
\end{equation} 
For notational convenience, we define $\chi_i^{\boldsymbol{\phi}}:=\partial_{\phi_i} \Phi,~ \chi_{ij}^{\boldsymbol{\phi}}:=\partial_{\phi_i}\partial_{\phi_j} \Phi$\,, and subsequent orders, which results in the cumulants, $\chi_i = \chi_i^{\boldsymbol{\phi}}|_{\boldsymbol{\phi}=\boldsymbol{0}},~ \chi_{ij} = \chi_{ij}^{\boldsymbol{\phi}}|_{\boldsymbol{\phi}=\boldsymbol{0}}$\,, respectively.
Additionally, we define $\langle G \rangle_{\boldsymbol{\phi}} := \sum_{\boldsymbol{g}}G(\boldsymbol{g}) P(\boldsymbol{g}, t)e^{\boldsymbol{\phi}^\top \boldsymbol{g}}\Big/\sum_{\boldsymbol{g}} P(\boldsymbol{g}, t) e^{\boldsymbol{\phi}^\top \boldsymbol{g}}$\,, which yields $\langle G \rangle_{\boldsymbol{\phi}}|_{\boldsymbol{\phi}=\boldsymbol{0}} = \langle G \rangle$.
Based on Price's equation and \eqref{eq:cumlant_generating}, along with the assumption that $P(\boldsymbol{g},t)$ does not induce strong interactions between sites and that the dynamics of higher-order cumulants are negligibly small, it is derived that the dynamics of first-order cumulants as $\dot{\chi}_i \simeq \sum_{j}\chi_{ij}\partial_{\chi_j}\langle F \rangle$ (ref.~\cite{neher2009competition}) The dynamics of arbitrary traits were expressed as a linear combination of first-order cumulant dynamics, $\frac{d\langle G \rangle}{dt} = \sum_i 
 \dot{\chi}_i\partial_{\chi_i}\langle G\rangle$. 
In the QLE framework, we assume that higher-order cumulants change faster than first-order cumulants, allowing the former to quickly reach equilibrium while the later are still evolving ($\dot{\chi}_{ij}=\dot{\chi}_{ijk}=\ldots=0$)\,.

\section*{Extension of QLE theory}\label{sec:generalization}
We begin by expressing the equation of motion for cumulants under any arbitrary genotype distribution. We then consider a scenario where the $K$-th order cumulants evolve dynamically, while higher-order cumulants (those of order $K+1$ and above) remain small and rapidly reach equilibria. 

To express the general form of cumulant dynamics, let $\mathcal{I, J, K}$ denote multi-indices over loci, e.g., $\mathcal{I} = (i_1, i_2, \ldots )$ where $i_1, i_2, \ldots$ are indices of loci.
Using the cumulant generating function and the relation
$
\partial_t \Phi(\boldsymbol{\phi}, t) 
= \langle F \rangle_{\boldsymbol{\phi}} - \langle F \rangle,
$
the dynamics of cumulants of arbitrary order can be expressed as
\begin{align}
    \begin{aligned}
        \dot{\chi}_{\mathcal{I}}
        &=\partial_t \partial_{\phi_{\mathcal{I}}}\Phi(\boldsymbol{\phi},t)|_{\boldsymbol{\phi}=\boldsymbol{0}}\\
        &= \partial_{\phi_{\mathcal{I}}}\langle F \rangle_{\boldsymbol{\phi}} |_{\boldsymbol{\phi}=\boldsymbol{0}}\,. 
    \end{aligned}
\end{align}

By denoting moments $\mu^{\boldsymbol{\phi}}_{\mathcal{I}} := e^{-\Phi}\partial_{\phi_{\mathcal{I}}}e^{\Phi}\, $ with $\mu_{\mathcal{I}} = \mu^{\boldsymbol{\phi}}_{\mathcal{I}}|_{\boldsymbol{\phi}=\boldsymbol{0}}$,
the last expression can be further written as:
\begin{align}
    \begin{aligned}
        \partial_{\phi_{\mathcal{I}}} \langle F \rangle_{\boldsymbol{\phi}}|_{\boldsymbol{\phi}=\boldsymbol{0}}
        &=\sum_{\mathcal{K}}
        \frac{\partial\langle F \rangle}{\partial\mu_\mathcal{K}} \left.\frac{\partial\mu_\mathcal{K}^{\boldsymbol{\phi}}}{\partial \phi_\mathcal{I}}\right|_{\boldsymbol{\phi}=\boldsymbol{0}}\\
        &=\sum_{\mathcal{J}, \mathcal{K}}
        \frac{\partial\langle F \rangle}{\partial\chi_\mathcal{J}} 
        \frac{\partial \chi_\mathcal{J}}{\partial\mu_\mathcal{K}} 
        \left.\frac{\partial\mu_\mathcal{K}^{\boldsymbol{\phi}}}{\partial \phi_\mathcal{I}}\right|_{\boldsymbol{\phi}=\boldsymbol{0}}
        \,.
    \end{aligned}
\end{align}
In the first equality, we use the fact that $\frac{\partial\langle F \rangle}{\partial\mu_L}$ is independent of the statistical variables. From the first to the second line, we convert from moment-based to cumulant-based expressions.

Therefore, by denoting $\boldsymbol{\chi}$ and $\boldsymbol{\nabla_\chi}$ as
$\boldsymbol{\chi} := ((\chi_i)_i, (\chi_{ij})_{i<j},\cdots)$ and $\boldsymbol{\nabla_{\boldsymbol{\chi}}} := ((\partial_{\chi_i})_i, (\partial_{\chi_{ij}})_{i<j},\cdots)$\,, 
the cumulant dynamics can be generally expressed as:
\begin{align}\label{eq:equation_generalized_QLE}
    \begin{aligned}
    \dot{\boldsymbol{\chi}} &~=    
    D(\boldsymbol{\chi})\boldsymbol{\nabla_{\chi}} \langle F \rangle\\
    D_{\mathcal{I,J}} &:=\sum_{\mathcal{K}}  
\frac{\partial \chi_\mathcal{J}}{\partial\mu_\mathcal{K}}         \left.\frac{\partial\mu_\mathcal{K}^{\boldsymbol{\phi}}}{\partial \phi_\mathcal{I}}\right|_{\boldsymbol{\phi}=\boldsymbol{0}}
    \end{aligned}
\end{align}
Explicit computations for $D_{\mathcal{I,J}}$ are provided in the \hyperref[sec:derive_D]{Supplementary Information (\hyperref[sec:genotype_dynamics_SI]{SI})}.
The matrix $D(\boldsymbol{\chi})$ is symmetric and positive definite, resulting from genotype competition and acting as a diffusion matrix in stochastic processes (see also \hyperref[sec:genotype_dynamics_SI]{SI}).
Alternatively, $D(\boldsymbol{\chi})$ can also be seen as a mobility matrix in fluid dynamics, as it links velocity (i.e., cumulant dynamics) to potential force (i.e., the gradient of average fitness) \cite{neher2011statistical, durlofsky1987dynamic}. 
\eqref{eq:equation_generalized_QLE} suggests that cumulants evolve along the gradient of average fitness through a matrix that serves as a geometric metric -- a picture that was also drawn in the QLE theory \cite{neher2011statistical} -- within this generalized setting.

Based on \eqref{eq:equation_generalized_QLE}, the extended QLE (exQLE) is derived by truncating the dynamics of cumulants up to order $K$.
In this setting, each row of $D(\boldsymbol{\chi})$, corresponding to cumulants up to order $K$, generally involves higher-order cumulants. For instance, the dynamics of first-order cumulants depend on second-, third-, and higher-order terms.
However, cumulant dynamics up to order $K$ depend only on a finite set of cumulants, determined by the fitness function. For example, in an additive fitness model ($K^* = 1$), the gradient of the first-order cumulant is constant, and those of higher-order cumulants vanish. Thus, only cumulants up to order $K + K^* = 2$ contribute to $D(\boldsymbol{\chi})$; higher-order terms ($>2$) have no effect on first-order dynamics.
Therefore, the equality in \eqref{eq:equation_generalized_QLE} holds as long as $D(\boldsymbol{\chi})$ includes cumulants up to order $K+K^*$, where $K^*$ is the highest-order cumulant contributing to the fitness function.
Similarly, if the goal is to understand the relationship between cumulant dynamics and the gradient of average fitness, and the fitness depends on cumulants up to order $K^*$, then considering cumulant dynamics up to order $K = K^*$ is necessary (note: the $D(\boldsymbol{\chi})$ matrix still needs to include cumulants up to order $K+K^*=2K^*$).
As we will discuss in the next section, the exQLE with $K=2$ under a pairwise fitness function ($K^*=2$) yields exact results for this reason. 

It is important to note that the expression in \eqref{eq:equation_generalized_QLE} is not self-contained, as the dynamics of cumulants up to order $K$ depend on higher-order cumulants up to order $K + K^*$, unless $\boldsymbol{\chi}$ includes cumulants of all orders. As a result, it is limited in describing cumulant dynamics over long time periods.  
To use this equation for simulating cumulant dynamics, the higher-order cumulants (beyond order $K$) must be specified externally (e.g., assumed to be zero or random) or described by using lower-order cumulants to make the equation self-contained.

\section*{Extended QLE: the $K=2$ case}\label{sec:second_order_dynamics}
To illustrate the extension of QLE theory, we consider the $K=2$ case, where cumulants up to the second order can dynamically evolve, while higher-order cumulants are small and rapidly reach equilibria.  
We assume a fitness function that depends on cumulants up to order $K^*>2$ and an arbitrary genotype distribution.  
Under these conditions, the dynamics of $\chi_{i}$ are given by:
\begin{align}\begin{aligned} \label{eq:extended_qle_1st}
    \dot \chi_i 
    &= \partial_{\phi_i}\langle F \rangle_{\boldsymbol{\phi}}|_{\boldsymbol{\phi}=\boldsymbol{0}} \\
    &\simeq 
    \sum_{k}\chi_{ik} \partial_{\chi_k} \langle F \rangle
    +
     \sum_{k<l}\chi_{ikl} \partial_{\chi_{kl}} \langle F \rangle\,.
\end{aligned}\end{align}
Similarly, the dynamics of $\chi_{ij}$ follow:  
\begin{align}\begin{aligned}\label{eq:extended_qle_2nd}
    \dot \chi_{ij}
    &= \partial_{\phi_{i}}\partial_{\phi_{j}} \langle F \rangle_{\boldsymbol{\phi}}|_{\boldsymbol{\phi}=\boldsymbol{0}}\\
    &\simeq 
    \sum_{k}\chi_{ijk}\partial_{\chi_k} \langle F \rangle
    + \sum_{k<l}(\chi_{ijkl}+\chi_{ik}\chi_{jl}+\chi_{il}\chi_{jk})\partial_{\chi_{kl}} \langle F \rangle\,.
\end{aligned}\end{align}
Details of the derivation \eqref{eq:extended_qle_2nd} are provided in \hyperref[sec:derivation_first_second_dynamis_SI]{SI}.
Note that $\dot{\chi}$ depends linearly on fitness parameters, which simplifies the inference of them as we discuss below.
If the average fitness function is characterized by cumulants up to seond-order, the relations in \eqref{eq:extended_qle_1st} and \eqref{eq:extended_qle_2nd} hold exactly. 
Under this extended QLE framework, the dynamics of any arbitrary trait $G$ is given as:
\begin{equation*}
    \frac{d\langle G \rangle}{dt} \simeq 
\sum_i \dot\chi_i \partial_{\chi_i}\langle G \rangle
+ 
\sum_{ij} \dot\chi_{ij} \partial_{\chi_{ij}}\langle G \rangle\,.
\end{equation*}

The dynamics effectively align with gradients of the average fitness surface and average traits.
The dynamics of cumulants are characterized by a symmetric matrix $D(\boldsymbol{\chi})$, such that 
\begin{align}\begin{aligned}\label{eq:definition_D_2nd_QLE}
\dot{\boldsymbol{\chi}} &\simeq D(\boldsymbol{\chi})\boldsymbol{\nabla_\chi} \langle F \rangle \\
D(\boldsymbol{\chi}) &= 
\begin{pmatrix}
\chi_{ik}
& \chi_{ikl} \\
\chi_{ijk}
& \chi_{ijkl}+\chi_{ik}\chi_{jl}+\chi_{il}\chi_{jk}
\end{pmatrix}\,.
\end{aligned}\end{align}
Here, the indices $i, j, k, l$ spans all combinations: the top-left block $\chi_{ik}$ spans rows and columns $1$ to $L$; the top-right spans rows $1$ to $L$, columns $L+1$ to $L(L+1)/2$; and the bottom-right spans rows and columns $L+1$ to $L(L+1)/2$.
Although the interpretation of the gradient of average fitness and the mobility matrix is absent, equations equivalent to \eqref{eq:extended_qle_1st} and \eqref{eq:extended_qle_2nd} can be found in ref.~\cite{dichio2023statistical}. 

\begin{figure}[t]
\includegraphics[width=\linewidth]{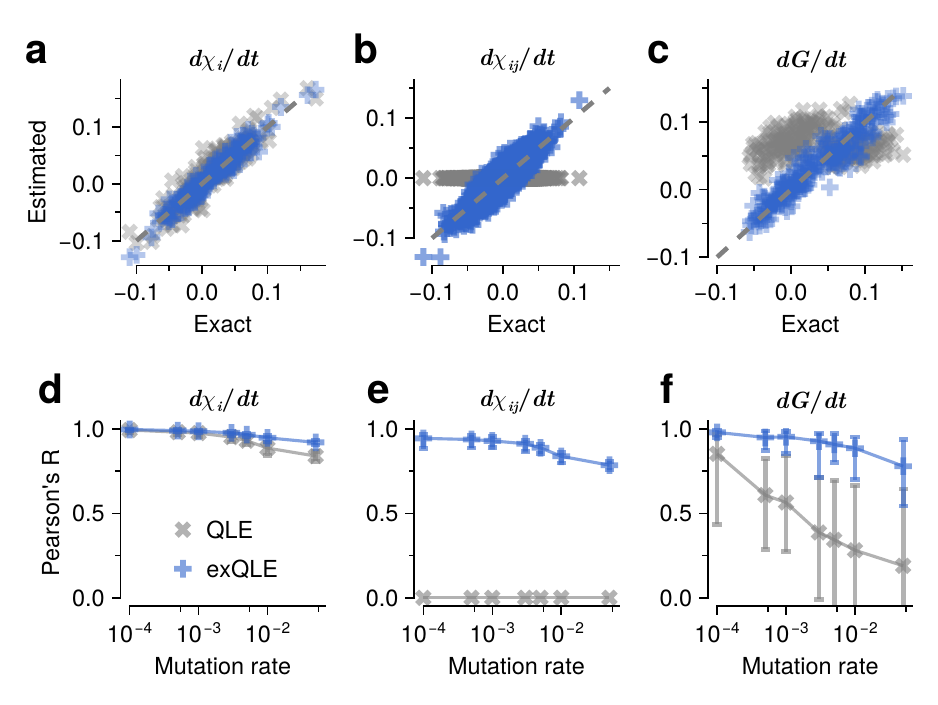}
\caption{\label{fig:rates_of_changes_in_1st_and_2nd_order_cumulant} 
\textbf{The accuracy of cumulant and trait dynamics by comparing estimated and exact values.}
To examine the accuracy of the exQLE framework, we consider
a fitness function that depends on interactions up to the fourth-order, providing a
non-trivial test case. Genetic sequences were simulated using the Wright–Fisher
(WF) process, incorporating mutation, recombination, and selection determined by
this fitness function. Simulation conditions are detailed at the \hyperref[sec:stochastic_process_SI]{SI}).
This presents the dynamics of first-order cumulants (\textbf{a}), second-order cumulants (\textbf{b}), and a random trait defined below (\textbf{c}), comparing results from exact calculations based on Price’s equation with those from the exQLE (with $K=2$) and QLE ($K=1$). 
Unlike the QLE case, exQLE closely match the exact dynamics for all cases. 
For the random trait, which is a pairwise function with parameters drawn from a normal distribution (see \hyperref[sec:stochastic_process_SI]{SI}), exQLE can accurately capture the dynamics, while QLE cannot. 
To systematically assess accuracy, we performed $10$ independent WF simulations and estimated the $R$ values across multiple mutation rates (\textbf{Fig.~\ref{fig:rates_of_changes_in_1st_and_2nd_order_cumulant}d-f}). 
Overall, exQLE consistently outperformed QLE, maintaining higher $R$ values across all mutation rates.
}
\end{figure}

To examine how the exQLE framework incorporates the second-order cumulant dynamics and enhances the descriptive capacity of population evolution, we numerically compared the cumulant dynamics derived in \eqref{eq:extended_qle_1st} and \eqref{eq:extended_qle_2nd} with the exact cumulant dynamics obtained from Price's equation.
As we discussed earlier, the expressions \eqref{eq:extended_qle_1st} and \eqref{eq:extended_qle_2nd} are exact when $K=K^*=2$, and the $D(\boldsymbol{\chi})$ includes cumulants up to the 4-th order, corresponding to $K+K^*$\,. 
Therefore, as a non-trivial example, we considered a fitness function incorporating third- and fourth-order interactions between sites (\textbf{Fig.~\ref{fig:rates_of_changes_in_1st_and_2nd_order_cumulant}}).  
The first- and second-order cumulant dynamics derived from exQLE closely correlate with those obtained from the exact calculations, whereas the original QLE framework fails to accurately describe the cumulant dynamics (\textbf{Fig.~\ref{fig:rates_of_changes_in_1st_and_2nd_order_cumulant}a-b}).
As an example of a more general trait, we evaluated the dynamics of a trait defined as a function of first- and second-order cumulants (its mathematical definition is provided in \hyperref[sec:stochastic_process_SI]{SI}). 
The results demonstrate that trait dynamics derived from exQLE align well with the exact dynamics, whereas QLE, which excludes second-order cumulant dynamics, does not (\textbf{Fig.~\ref{fig:rates_of_changes_in_1st_and_2nd_order_cumulant}c}). 

To assess the accuracy of exQLE in capturing cumulant and trait dynamics, we analyzed the correlation between the estimated $\dot{\chi}_i, \dot{\chi}_{ij}$ and random traits across multiple mutation rates. 
Increasing mutation rates lead to a greater number of genotypes, effectively enhancing the presence of finite higher-order cumulants. 
The Pearson's $R$ values comparing those from the exact results and estimated ones based on the exQLE framework achieve consistently higher values than those from the QLE method across various mutations (\textbf{Fig.~\ref{fig:rates_of_changes_in_1st_and_2nd_order_cumulant}d-f}). 

Naive computation of dynamics under higher-order fitness is computationally challenging; therefore, we also propose an efficient method in \hyperref[sec:efficient_computation_SI]{SI}.

\section*{Inferring fitness parameters}\label{sec:inference}
We now illustrate how to infer fitness parameters $\boldsymbol{s} = ((s_i)_i, (s_{ij})_{i<j}, \ldots)$ from cumulant dynamics. As discussed at the end of the section, the core idea parallels the derivation of the recently proposed marginal path-likelihood (MPL) method \cite{sohail2021mpl, sohail2022inferring}.
A more mathematically explicit discussion is also provided in \hyperref[sec:derive_MLE_for_cumulants_SI]{SI}. 

So far, our analysis has assumed deterministic dynamics. However, under finite population size $N$, stochasticity must be incorporated. To account for this effect, the cumulant dynamics can be represented as a Langevin equation:
\begin{equation}\label{eq:Langevin}
\dot{\boldsymbol{\chi}} = D(\boldsymbol{\chi})\boldsymbol{\nabla_\chi} \langle F \rangle + \sqrt{D(\boldsymbol{\chi})/N}\, \boldsymbol{\eta}(t)\,,
\end{equation}
where $\boldsymbol{\eta}(t)$ is white Gaussian noise, satisfying with $\langle \boldsymbol{\eta} \rangle=\boldsymbol{0}$ and $\langle \eta_{\mathcal{J}}(t) \eta_{\mathcal{K}}(t') \rangle= \delta_{\mathcal{J,K}}\delta(t-t')$\,. This form of noise arises under the WF process and has been used in the prior work \cite{neher2011statistical, sohail2021mpl, sohail2022inferring}.


Due to stochasticity $\boldsymbol{\eta}(t)$, the system can trace varied cumulant trajectories over time.
The probability distribution of the cumulant trajectories over time, $P((\boldsymbol{\chi}(t_k))_{k=0}^{K}|\boldsymbol{s})$, is obtained from the Langevin equation. The most probable parameter $\hat{\boldsymbol{s}}$ maximizes this likelihood and equivalently, minimize the action, the time integral of the squared noise term (see \hyperref[sec:genotype_dynamics_SI]{SI}). Therefore, the solution is given by:
\begin{align}
    \begin{aligned}
        \hat{\boldsymbol{s}} &=
        \underset{\boldsymbol{s}}{\arg\max}\,\,P((\boldsymbol{\chi}(t_k))_{k=0}^{K}|\boldsymbol{s})\\
        &= \left(\sum_{k=0}^{K} \Delta t_{k} \tilde{D}(\boldsymbol{\chi}(t_k))\right)^{-1} 
        \sum_{k=0}^{K} \boldsymbol{\Delta \chi}(t_k)\,,
    \end{aligned}
\end{align}
where $\Delta t_k := t_{k+1}- t_k$, $\boldsymbol{\Delta \chi}(t_k) := \boldsymbol{\chi}(t_{k+1}) - \boldsymbol{\chi}(t_{k})$, and the matrix $\tilde{D}(\boldsymbol{\chi})$ is defined such that
\begin{equation}\label{eq:D_and_tilde_D}
\tilde{D}(\boldsymbol{\chi}) \boldsymbol{s}:=D(\boldsymbol{\chi})\boldsymbol{\nabla_\chi}\langle F \rangle  \,. 
\end{equation}
The simple closed-form inference arises from the linearity of the dynamics in the fitness parameters (see \hyperref[sec:derive_MLE_for_cumulants_SI]{SI}).
This path-likelihood maximization framework provides a principled basis for inference, with statistical integrals arising naturally in the solution.

As an example, in the $K=2$ case, the maximum likelihood equation can be used to infer $\boldsymbol{s}$ under WF process with mutation and recombination. By reintroducing the expected cumulant changes due to mutation and recombination, as computed in prior work \cite{neher2011statistical, dichio2023statistical}, the solution becomes:
\begin{align}\label{eq:cumulant_dynamics_1st_2nd_with_mutation_recombination}
    \begin{aligned}
        \hat{\boldsymbol{s}}&= \left( \sum_{k=0}^K \Delta t_k \tilde{D}(\boldsymbol{\chi}(t_k)) \right)^{-1}\\
        &\qquad \times \sum_{k=0}^K \left[
        \begin{pmatrix}
        \Delta\chi_i(t_k) \\
        \Delta \chi_{ij}(t_k)
        \end{pmatrix}
        + 
        \Delta t_k
        \begin{pmatrix}
            2\mu \chi_i(t_k)\\
            (4\mu + r c_{ij})\chi_{ij}(t_k)
        \end{pmatrix}\right] \,,
    \end{aligned}
\end{align} 
with
\begin{equation}\label{eq:D_tilde_matrix}
\tilde{D}(\boldsymbol{\chi}) = \begin{pmatrix}
            {\small \chi_{ik}}
            & {\small \chi_{ikl} + \chi_{ik}\chi_{l} + \chi_{il}\chi_{k}} \\
            {\small \chi_{ijk}}
            & {\small \chi_{ijkl} + \chi_{ik}\chi_{jl} + \chi_{il}\chi_{jk} + \chi_{ijk}\chi_l + \chi_{ijl}\chi_k}
            \end{pmatrix},
\end{equation}
which is derived from the definition of $\tilde{D}$ in \eqref{eq:D_and_tilde_D} and the matrix $D(\boldsymbol{\chi})$ in \eqref{eq:definition_D_2nd_QLE}\,.

This expression, \eqref{eq:cumulant_dynamics_1st_2nd_with_mutation_recombination} and \eqref{eq:D_tilde_matrix}, matches with the one derived from MPL upon transforming from the $0/1$ basis and moment representation to the $-/+$ basis and cumulants \cite{sohail2021mpl, sohail2022inferring, shimagaki2025efficient}. 
Besides mathematical conventions, exQLE and MPL differ in that exQLE employs a Langevin equation for cumulant dynamics for arbitrary order $K$, while MPL considers the WF process under the diffusion limit. 

The cumulant-based approach allows for systematic approximations of the equation of motion by reducing the effects of higher-order cumulants in an order-by-order manner. For example, given the expression 
\eqref{eq:cumulant_dynamics_1st_2nd_with_mutation_recombination}, it is relevant to consider the scenario where cumulants beyond the second order are negligible. In this case, $D(\boldsymbol{\chi})$ matrix depends only on second-order cumulants, leading to 
\begin{align}\begin{aligned}\label{eq:gaussian_closure_exQLE}
\dot\chi_i &\simeq \sum_{k}s_k \chi_{ik} + \sum_{k<l}s_{kl}\chi_{ik}\chi_l - 2\mu\chi_i \\
\dot\chi_{ij} &\simeq \sum_{k<l} s_{kl}
(\chi_{il}\chi_{jk}+ \chi_{jl}\chi_{ik}) - (4\mu + r c_{ij})\chi_{ij}\,. 
\end{aligned}\end{align}
When considering only up to second-order cumulants, the dependence of second-order cumulant dynamics on additive selection disappears, as also observed in the previous multi-locus QLE study \cite{neher2011statistical}. 
The absence of additive fitness dependence in second-order cumulant dynamics, which ultimately enables a closed-form expression for epistasis \cite{neher2011statistical, dichio2023statistical}, arises because epistatic effects are coupled with third-order cumulants. By ignoring third-order cumulant dynamics, cumulant evolution becomes effectively independent of additive fitness effects.
Suppose we assume steady-state conditions in $\chi_{ij}$ and further approximate $\sum_{k<l}s_{kl}\chi_{il}\chi_{jk} \simeq s_{ij}\chi_{ii}\chi_{jj}$. In this case, we recover the identical solution as the QLE epistasis model with Gaussian closure (GC) scheme, which drops more than second-order cumulants \cite{mauri2021gaussian, zeng2021inferring}, given as: $s_{ij} = (4\mu + rc_{ij})\ \chi_{ij} / \chi_{ii}\chi_{jj}\,.$

To evaluate how accurately the exQLE inference framework, represented by \eqref{eq:cumulant_dynamics_1st_2nd_with_mutation_recombination}, and the exQLE with GC scheme (exQLEGC), represented by \eqref{eq:gaussian_closure_exQLE}, can infer fitness parameters including selection and epistatic coefficients, we performed 100 independent WF simulations under a pairwise fitness function across multiple mutation rates. This allows us to systematically assess the correlation between inferred coefficients and ground-truth fitness parameters. \textbf{Fig.~\ref{fig:SI_Pearson_HI}} also illustrates the results from the $K^*=4$ case. 

\begin{figure}[t]
\centering
\includegraphics[width=0.8\linewidth]{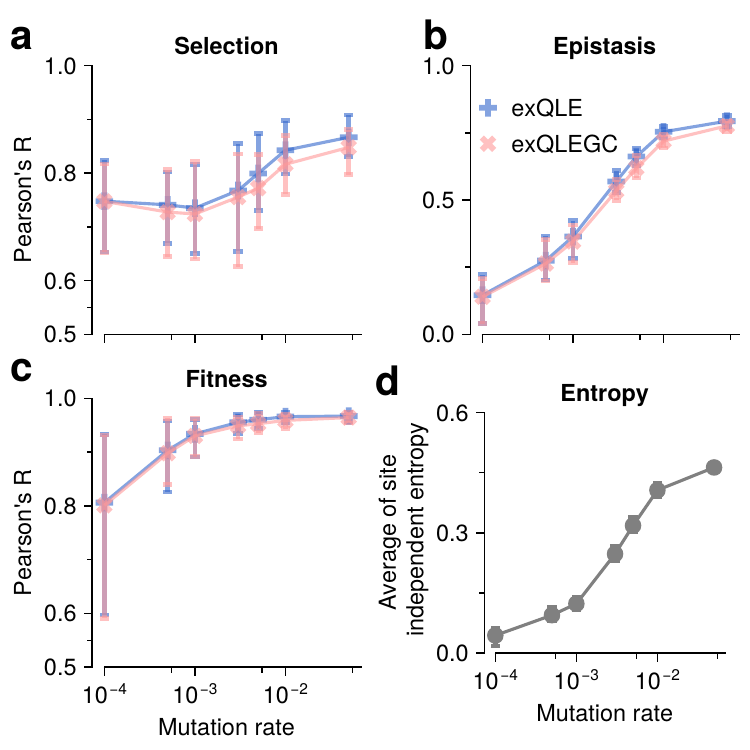}
\caption{\label{fig:accuracy_PW} 
Pearson’s $R$ values compare inferred and ground-truth fitness for selection coefficients (\textbf{a}), epistatic coefficients (\textbf{b}), and overall fitness (\textbf{c}), shown for $K^*=2$ (with $K^*=4$ in \textbf{Fig.~\ref{fig:SI_dynamics_exQLEGC}}). The recombination rate was fixed at $r=3\times10^{-3}$ per generation per site, as recombination minimally affects inference accuracy, especially under exQLE. Inferred coefficients were obtained from temporal genetic sequences generated with the WF process (including recombination, mutation, and selection) based on the cumulant dynamics expressions \eqref{eq:cumulant_dynamics_1st_2nd_with_mutation_recombination} (or \eqref{eq:gaussian_closure_exQLE} for QLEGC).
Inference of selection coefficients remains consistently accurate across mutation rates, whereas inference of epistatic coefficients strongly depends on mutation rates. This dependency reflects the effect of genetic diversity: higher mutation rates increase diversity, improving estimation of higher-order cumulants. Genetic diversity is quantified by the entropy profile, defined as the entropy of independent site frequency $\nu_i=(1+\chi_i)/2$:
$S=-\frac{1}{L}\sum_{i=1}^L \langle \nu_i(t)\log(\nu_i(t)) \rangle_t$,
averaged every 50 generations between 1,500 and 2,000. The entropy profile closely parallels the $R$-value profile for epistatic inference, confirming the link between mutation-driven diversity and improved accuracy.
}
\end{figure}

Overall, exQLE and exQLEGC achieve similar inference accuracy (accuracy of the forward simulation using exQLEGC can be found in \textbf{Fig.~\ref{fig:SI_dynamics_exQLEGC}}). For selection coefficients, the average Pearson's $R$ values range from $0.74$ to $0.87$, with the lowest $R$ value occurring at smaller mutation rates and the highest $R$ values at the highest mutation rate ($\mu=0.05$) (\textbf{Fig.~\ref{fig:accuracy_PW}a}).
For epistatic inference, mutation rates noticeably influence $R$ values, with the lowest $R$ value of $0.15$ occurring at the lowest mutation rate. 
Higher mutation rates yielded higher $R$ values, up to a maximum of $0.79$ (\textbf{Fig.~\ref{fig:accuracy_PW}b}). 
Genetic diversity is measured by the average entropy across all sites, assuming that the sites are independent of one another, and varies with mutation rates in a similar manner to the Pearson's $R$ values for inferred epistasis. 
This suggests that increased mutation rates diversify the population, effectively increasing the number of distinct genotypes, enhancing the precision of cumulant estimation, and resulting in higher $R$ values for the epistasis inference (\textbf{Fig.~\ref{fig:accuracy_PW}d}). 
The inferred fitness values, obtained from inferred selection and epistatic coefficients as well as genetic sequences, also increase with mutation rates, though they exceed $0.81$ with a maximum value of $0.97$ at the highest mutation rate. 

\section*{Discussion}\label{sec:discussion}
Quantitatively understanding how fundamental evolutionary forces, such as selection, mutation, and recombination, shape the evolution of populations and phenotypic traits, is a central question in evolutionary biology \cite{lande1983measurement, walsh2018evolution, nourmohammad2013universality, nourmohammad2017adaptive}. 
While considering multilocus effects is crucial for capturing the complexity of genetic evolution, and growing evidence shows that most phenotypic traits are governed by many alleles across different loci \cite{chou2011diminishing, khan2011negative, kryazhimskiy2014global, olson2014comprehensive, starr2018pervasive, park2022epistatic, bakerlee2022idiosyncratic, sailer2017high, poelwijk2019learning, phillips2021binding, moulana2023landscape, phillips2023hierarchical, yang2019higher, buda2023pervasive, schulz2025epistatic}, many foundational results were based on single-locus or two-locus models \cite{wright1931evolution, wright1937distribution, crow1965evolution, kimura1964diffusion, kimura1965attainment, ohta1969linkage}. 

Existing studies of genetic evolution in multilocus populations often assume that the non-random association between alleles in different loci, known as linkage disequilibrium (LD), rapidly vanishes and is negligibly small \cite{barton1989evolutionary, kirkpatrick2002general, lynch1998genetics, wright1937distribution}. Although these studies provide valuable insights into genetic evolution, they do not apply to collective evolution across different sites, which involves complex and rich phenomena \cite{smith1974hitch, felsenstein1974evolutionary, barton2000genetic, hill1966effect, good2014genetic, good2022linkage}.

Quasi-linkage equilibrium (QLE), where the LD is present but is weak and rapidly converges to the equilibrium state \cite{kimura1965attainment, crow2017introduction, barton1991natural, turelli1994genetic, nagylaki1993evolution, barton1995general,  kirkpatrick2002general, barton2005evolution, neher2011statistical}, allows for exploring collective allele evolution while simplifying mathematical structures. 
However, the existing QLE theories have limitations. Populations can reach the QLE phase only when the recombination rate is much larger than the selection \cite{barton2005evolution}, and epistasis is also smaller than the recombination rate \cite{neher2009competition}, which significantly restricts the application of the QLE phase. 
In the inter-host evolution of SARS-CoV-2, selective pressures are likely strong, but the recombination rate is effectively zero, leading to significant LD \cite{lee2025inferring}. Although the human immunodeficiency virus exhibits a higher recombination rate \cite{zanini2015population, romero2024elevated}, selective pressure can be significant, and often substantial LD persists over many years \cite{zanini2015population, sohail2021mpl, shimagaki2025parallel, shimagaki2023bezier}. 
Additionally, the recent progress in high-throughput deep mutational scanning to measure functional effects revealed the prevalence of widespread epistatic epistasis in viral and bacterial population in wild, and bacterial populations \cite{otwinowski2018inferring, starr2020deep, moulana2022compensatory, moulana2023landscape}.
Collectively, this evidence suggests that the QLE assumption violates conditions where LDs or higher-order cumulants of alleles across loci are unlikely to converge to equilibrium rapidly, and the timescales of higher-order cumulants and individual allele frequencies are not well separated. 

Here, we present an extension of QLE theory, exQLE, which generalizes QLE to allow cumulants up to any order $K$ to evolve dynamically.
To demonstrate this, we first expressed the cumulant dynamics for arbitrary orders and arbitrary genotype distributions (corresponding to \eqref{eq:equation_generalized_QLE} ), providing a geometric interpretation of the dynamics and insights into when the expression can be exact.
The resulting matrix $D(\boldsymbol{\chi})$ servers as a geometric metric, and its general form is detailed in the \hyperref[sec:derive_D]{Supplementary Information}.
The exQLE formulation naturally arises from \eqref{eq:equation_generalized_QLE} under conditions or genotype distributions where cumulants of order greater than $K$ evolve rapidly toward equilibria, while the $K$-th order and lower cumulants remain dynamically changing.
As an example of exQLE, we investigated the case $K=2$, focusing on fitness functions whose averages depend on cumulants up to order $K^*=4$ (noting that the $K^*=2$ case is trivially exact). Under this condition, the $K=2$ exQLE framework can accurately demonstrate the cumulant dynamics, a capability that the standard QLE framework lacks.
The derived cumulant dynamics also enable inference of fitness parameters by maximizing the likelihood of observed cumulant trajectories over evolutionary time, as in the marginal path-likelihood (MPL) method\cite{sohail2021mpl, sohail2022inferring, gao2025binary, shimagaki2025efficient}.

Since the equations for cumulant dynamics are expressed as combinations of cumulants, their complexity can be systematically reduced by suppressing cumulants in a stratified, order-by-order manner.
To achieve this, we employed an alternative Gaussian closure scheme, in which cumulants beyond the second order are assumed to be absent. 
The resulting novel family of inference methods accurately estimates selection and epistatic coefficients.

We could not fully explore the equilibrium distribution in the exQLE framework, in this work. Similar to the prior work \cite{neher2011statistical}, an explicit expression for the equilibrium distribution $Q(\boldsymbol{\chi})$ could be derived from the forward Kolmogorov equation, which is characterized by $K$-th order cumulant dynamics, such that $\partial_t Q (\boldsymbol{\chi},t) = \boldsymbol{\nabla_\chi}^\top \boldsymbol{j}(\boldsymbol{\chi},t)$\,, where $\boldsymbol{j}(\boldsymbol{\chi},t)$ is the probability current. 
The equilibrium distribution can be obtained from the condition
$\boldsymbol{j}(\boldsymbol{\chi})=\boldsymbol{0}$. 
For example, when $K=2$, the equilibrium distribution includes the interactions between first-order cumulants in an exponential function. 
Similar to the equilibrium distribution under QLE \cite{neher2011statistical}, the equilibrium distribution under exQLE also features an exponential form and an entropic term. 
Although, the general form of the equilibrium distribution remains conjectural, it likely depends on higher-order epistatic interactions between all possible combinations of cumulants up to order $K$, shaped by cumulants, mutation rates, and recombination rates, and appearing within the exponential function.

\begin{acknowledgements}
The work of K.S.S.~and J.P.B.~reported in this publication was supported by the National Institute of General Medical Sciences of the National Institutes of Health under Award Number R35GM138233. 
The work of M.P. was supported by the European Union (ERC, CHORAL, project number 101039794). Views and opinions expressed are however those of the authors only and do not necessarily reflect those of the European Union or the European Research Council. Neither the European Union nor the granting authority can be held responsible for them.
\end{acknowledgements}



%

%% file: SI.tex
\renewcommand{\d}{\mathrm{d}}

\clearpage
\newpage
{\fontsize{20pt}{20}\bfseries\centering
\bf{Supplementary Information}}

\numberwithin{equation}{section}
\renewcommand{\theequation}{S\arabic{equation}}
\setcounter{equation}{0}
\renewcommand{\thefigure}{S\arabic{figure}}
\setcounter{figure}{0}



\section*{Stochastic Processes in Population genetics}\phantomsection\label{sec:stochastic_process_SI}
Following the convention of previous studies \cite{neher2011statistical, zeng2020inferring}, we represent genetic sequences as $\boldsymbol{g} \in \{-1, +1\}^{L}$, where $L$ is the sequence length. This $-/+$ encoding is commonly used in the physics literature. For a genotype $a \in \{1, \ldots, M\}$ with $M = 2^L$, and corresponding sequence $\boldsymbol{g}^a$, the fitness is defined as $F^a = F(\boldsymbol{g}^a)$. We consider the following general form for the fitness function:
\begin{equation}
    F(\boldsymbol{g}) = \bar{F} + \sum_i s_i g_i + \sum_{i<j}s_{ij}g_i g_j + \cdots\,,
\end{equation}
where the fitness parameters $s_i$ and $s_{ij}$ represent selection coefficients and pairwise epistatic interactions, respectively.
Interactions beyond pairwise, such as $s_{ijk}, s_{ijkl}, \ldots$, can also occur, and we refer to these as higher-order epistatic interactions.
For simplicity, we denote fitness parameters generically as $s_e$, where the index $e$ may represent a single-site effect $(i)$, pairwise $(i,j)$, three-way $(i,j,k)$, four-way $(i,j,k,l)$, and so on.

To model the evolution of genetic sequences, we employed the Wright–Fisher (WF) process, a foundational stochastic model in population genetics that captures reproduction dynamics \cite{wright1931evolution}.
The WF process describes the evolution of a population with fixed size $N$. At time $t_k$ (the parent generation), the population is represented by genotype counts $(n_1(t_k), \ldots, n_M(t_k))^\top = (n_a(t_k))_a^M$, which give rise to the next generation at time $t_{k+1}$, denoted $(n_a(t_{k+1}))_a^M$.

Demographic noise, or genetic drift, arises from stochastic sampling and is on the order of $\mathcal{O}(1/N)$. It can be quantified through fluctuations in genotype frequencies, defined as $\nu_a = n_a/N$.
Mathematically, the WF process is a discrete-time multinomial process. Additional evolutionary forces, including selection (with fitness $F(\boldsymbol{g})$), mutation, and recombination, can be incorporated. Mutation and recombination occur at rates $\mu$ and $r$ per site per generation, respectively.

Let $\boldsymbol{\nu} = \boldsymbol{n}/N$ denote the genotype frequency vector, $p_{a}(\boldsymbol{\nu} \mid F, \mu, r)$ denote the probability that genotype $a$ is selected, given the current genotype frequencies $\boldsymbol{\nu}$, fitness function $F(\boldsymbol{g})$, and evolutionary forces such as mutation and recombination (defined below).
Then, the WF process can be expressed as:

\begin{equation}
    p((\boldsymbol{\nu}(t_k))_{k=0}^K|F, \mu, r, N)
    =\prod_{k=0}^{K-1} p(\boldsymbol{\nu}(t_{k+1})|\boldsymbol{\nu}(t_{k});F, \mu, r, N)  \,,
\end{equation}
whre 
\begin{equation}
    p(\boldsymbol{\nu}(t_{k+1})|\boldsymbol{\nu}(t_{k})\ ;F, \mu, r, N) 
    = N!\displaystyle\prod_{a}\frac{
    p_{a}( \boldsymbol{\nu}(t_k)| F, \mu, r)^{N\nu_a(t_{k+1})} 
    }{[N\nu_a(t_{k+1})]\ !}\,.
\end{equation}

Let $y_a(\boldsymbol{\nu}; r)$ denote the probability that recombination events produce genotype $a$, which can be expressed as:
\begin{equation}\label{eq:recombination_prob}
    y_a(\boldsymbol{\nu}; r) = (1-r)^{L-1}\nu_a + \left( 1 - (1-r)^{L-1}\right)\sum_{b,c}R_{a|b,c}\nu_b \nu_c\,.
\end{equation}

The selection probability is then given by: 
\begin{equation}\label{eq:selection_prob}
    p_{a}( \boldsymbol{\nu}| F, \mu, r)
    = \frac{y_a(\boldsymbol{\nu}; r) F_a + \mu\sum_{b;\, d_{ab}=1}[y_b(\boldsymbol{\nu}; r) F_b - y_a(\boldsymbol{\nu}; r) F_a ]}{\sum_b y_b(\boldsymbol{\nu}; r) F_b}\,.
\end{equation}
In this expression, $d_{ab}$ denotes the Hamming distance between genotypes $a$ and $b$.
When the mutation rate $\mu$ is low, at most one mutation is expected per individual per generation. As a result, the contribution to the mutation flux in the numerator of \eqref{eq:selection_prob} comes only from genotype pairs that differ by a single mutation (i.e., $d_{ab} = 1$).

\subsection*{Details of simulation conditions}\phantomsection\label{sec:details_of_simulation_SI}
To examine a non-trivial scenario, we considered a higher-order fitness function defined as
\begin{equation}
    F(\boldsymbol{g}) 
= \bar{F}+\sum_i s_i g_i + \sum_{i<j}s_{ij}g_i g_j 
+ \sum_{i<j<k}s_{ijk}g_i g_j g_k+ \sum_{i<j<k<l}s_{ijkl}g_i g_j g_k g_l\,.
\end{equation}
where the fitness coefficients $s_e \in \{-0.03, 0, 0.03\}$ for indices $e = i, (i,j), (i,j,k), (i,j,k,l)$. The number of nonzero coefficients $s_e$ was kept at $\mathcal{O}(L)$ across orders one through four, with $L$ denoting the sequence length.
In Fig.~1, to assess the influence of higher-order cumulants on trait dynamics, we also considered random traits defined as
\begin{equation}
    G^{\mathrm{Rand}}(\boldsymbol{g}) = \sum_i a_i g_i + \sum_{i<j}a_{ij}g_i g_j\,,
\end{equation}
where $a_i \sim \mathcal{N}(0, 1/L)$ and $a_{ij} \sim \mathcal{N}(0, 2/L(L-1))$ for all $i, j$. These coefficients were independently sampled.

The simulation conditions of \textbf{Fig.~\ref{fig:rates_of_changes_in_1st_and_2nd_order_cumulant}} are as follows. Fitness parameters were drawn from the set $\{-0.03, 0, 0.03\}$, while maintaining the number of nonzero parameters at $O(L)$ for each order, where $L = 100$ is the sequence length. The recombination rate was fixed at $r = 3 \times 10^{-3}$ per site per generation, as the results were robust across different recombination rates. In contrast, the simulation outcomes depend on the mutation rate, which we varied between $10^{-4}$ and $0.05$ per site per generation. The population size was set to $N = 10^3$. As we observed no significant variation in cumulant dynamics after $10^3$ generations, we sampled genetic sequences every $200$ generations between $10^3$ and $2\times10^3$ generations.

\section*{Computation of $D_{\mathcal{I,J}}$ matrix}
We now provide a more explicit expression for $D_{\mathcal{I,J}}$. 
As we noted in the main text, $\mathcal{I, J, K}$  are multi-indices over loci, and cumulants and moments of arbitrary order are defined as $\chi_{\mathcal{I}}^\phi=\partial_{\phi_{\mathcal{I}}}\Phi$\, and $\mu_{\mathcal{I}}^\phi = e^{-\Phi}\partial_{\phi_{\mathcal{I}}}e^{\Phi}\,, $  respectively. 
The general form of cumulant dynamics is given as
\begin{align}\label{eq:def_D}
    \begin{aligned}
        \partial_{\phi_{\mathcal{I}}} \langle F \rangle_{\boldsymbol{\phi}}|_{\boldsymbol{\phi}=\boldsymbol{0}}
        &=
        \sum_{\mathcal{J}} D_{\mathcal{I,J}}
        \frac{\partial\langle F \rangle}{\partial\chi_\mathcal{J}} 
        \\
        D_{\mathcal{I,J}} &= \sum_{\mathcal{K}} \frac{\partial \chi_\mathcal{J}}{\partial\mu_\mathcal{K}} 
        \left.\frac{\partial\mu_\mathcal{K}^{\phi}}{\partial \phi_\mathcal{I}}\right|_{\boldsymbol{\phi}=\boldsymbol{0}}\,.
    \end{aligned}
\end{align}

We express moments $\mu_{\mathcal{K}}$ in terms of cumulants since derivatives of cumulants with respect to $\phi$ are more tractable. Specifically, moments can be written as:
\begin{equation}
    \mu_{\mathcal{K}} = \sum_{\pi \in \mathcal{P}(\mathcal{K})}\prod_{B\in\pi}\chi_B\qquad. 
\end{equation}
where $\mathcal{P}(\mathcal{K})$ denotes all partitions of the index set $\mathcal{K}$. For example, if $\mathcal{K} = \{k_1\}$, then $\mathcal{P}(\mathcal{K}) = \{\{k_1\}\}$. For $\mathcal{K} = \{k_1, k_2\}$,
$
\mathcal{P}(\mathcal{K}) = \big\{ \{k_1, k_2\}, \{\{k_1\}, \{k_2\}\} \big\},
$
and so on.

Therefore, 
\begin{equation}\label{eq:mu_over_phi}
    \left.
    \frac{\partial\mu_{\mathcal{K}}^{\phi}}{\partial \phi_{\mathcal{I}}}
    \right|_{\boldsymbol{\phi} = \boldsymbol{0}}
    = \sum_{\pi \in \mathcal{P}(\mathcal{K})} \left[ \partial_{\phi_{\mathcal{I}}}\prod_{B\in\pi}\chi_{B}^{\phi}\right]_{\boldsymbol{\phi} = \boldsymbol{0}}\,. 
\end{equation}

To relate cumulants to moments, we use the Faà di Bruno formula:
\begin{equation}
    \chi_{\mathcal{J}} = \sum_{\pi \in \mathcal{P}(\mathcal{J})} (-1)^{|\pi|-1}(|\pi|-1)!\prod_{B \in \pi}\mu_B\,,
\end{equation}
from which, we obtain the derivative of cumulants with respect to moments: 
\begin{equation}\label{eq:chi_over_mu}
    \frac{\partial \chi_{\mathcal{J}}}{\partial \mu_{\mathcal{K}}} = \sum_{\pi \in \mathcal{P}(\mathcal{J})} (-1)^{|\pi|-1}(|\pi|-1)! \frac{\partial}{\partial \mu_{\mathcal{K}}}\prod_{B \in \pi}\mu_B\, .
\end{equation}
Thus, combining \eqref{eq:chi_over_mu} and \eqref{eq:mu_over_phi}, we obtain the explicit expression of $D_{\mathcal{I,J}}$ in \eqref{eq:def_D}\,.
Below, we demonstrate this for specific cases under a pairwise fitness function. 
\subsubsection*{\underline{ $K=1$ case}}
Let us consider a simple $\mathcal{I}=\{ i \}\,, \mathcal{J}=\{ j \}$ case. 
The only partition for $\mathcal{K} = \{j\}$ is $\mathcal{P}(\mathcal{K}) = \{\{j\}\}$, and we obtain:
$$
\frac{\partial \chi_{j}}{\partial \mu_{\mathcal{K}}} =  \delta_{\{j\}, \mathcal{K}}\,,$$
where $\delta_{\mathcal{K}, \mathcal{L}}$ returns 1 if $\mathcal{K}= \mathcal{L}$, otherwise returns 0. 
Thus, the $D$ matrix reduces to:
$$
D_{i,j} = \sum_{\mathcal{K}}\delta_{\{ j\}\mathcal{K}}
\sum_{\pi \in \mathcal{P}(\mathcal{K})} \left[ \partial_{\phi_i}\prod_{B\in\pi}\chi_{B}^{\phi} \right]_{\boldsymbol{\phi} = \boldsymbol{0}}= 
\sum_{\mathcal{K}}\delta_{\{ j\}\mathcal{K}}
\partial_{\phi_i}\chi_j^{\phi}|_{\boldsymbol{\phi}=\boldsymbol{0}}
=
\chi_{ij}\, .
$$
where only $\mathcal{K} = \{j\}$ contributes. Therefore, this result is consistent with the $K=1$ case.

\subsubsection*{\underline{$K=2$ case}}
Since the expression for $D_{ij}$ matches the $K=1$ case, we consider three additional cases:
$\mathcal{I}=\{i,j\},\, \mathcal{J}=\{ k \}$;
$\mathcal{I}=\{ i \},\, \mathcal{J}=\{k,l\}$; and 
$\mathcal{I}=\{i,j\},\, \mathcal{J}=\{ k, l\}$\,.

For $\mathcal{I}=\{ i, j \},\, \mathcal{J}=\{k\}$, 
we have seen that $\frac{\partial \chi_{k}}{\partial \mu_{\mathcal{K}}}
= \delta_{j, k}\,.$ from the example in $K=1$ case. Therefore, we have
$$
D_{ij,k} = \sum_{\mathcal{K}}\delta_{\{k\},\mathcal{K}}
\left[ \partial_{\phi_i}\partial_{\phi_j}\chi_{k}^{\phi} \right]_{\boldsymbol{\phi} = \boldsymbol{0}} = \chi_{ijk}\, . 
$$

For $\mathcal{I}=\{ i \},\, \mathcal{J}=\{k,l\}$, the cumulant-moment relation yields, 
$$
\frac{\partial \chi_{kl}}{\partial \mu_{\mathcal{K}}} 
= -\chi_l\delta_{\{k\},\mathcal{K}}  
-\chi_k\delta_{\{l\},\mathcal{K}} 
+ \delta_{\{k,l\},\mathcal{K}} \,.
$$

Therefore, 
\begin{align*}
    \begin{aligned}
D_{i, kl} &= \sum_{\mathcal{K}}\left(
-\chi_l\delta_{\{k\},\mathcal{K}}  
-\chi_k\delta_{\{l\},\mathcal{K}} 
+ \delta_{\{k,l\},\mathcal{K}}
\right)
\left[\partial_{\phi_i}
\sum_{\pi \in \mathcal{P}(\mathcal{K})}\prod_{B\in\pi}\chi_{B}^{\phi}
\right]_{\boldsymbol{\phi} = \boldsymbol{0}}
\\
&= 
-\chi_l\chi_{ik}
-\chi_k\chi_{il}
+ 
\left[\partial_{\phi_i}
(\chi_{kl}^{\phi}+\chi_{k}^{\phi}\chi_{l}^{\phi})
\right]_{\boldsymbol{\phi} = \boldsymbol{0}}\\
&= 
-\chi_l\chi_{ik}
-\chi_k\chi_{il}
+ \chi_{ikl}
+\chi_l\chi_{ik}
+\chi_k\chi_{il}
= \chi_{ikl}\, .
\end{aligned}
\end{align*}

For $\mathcal{I}=\{i,j\},\, \mathcal{J}=\{ k,l\}$ case, 
\begin{align*}
    \begin{aligned}
D_{ij, kl} &= \sum_{\mathcal{K}}\left(
-\chi_l\delta_{\{k\},\mathcal{K}}  
-\chi_k\delta_{\{l\},\mathcal{K}} 
+ \delta_{\{k,l\},\mathcal{K}}
\right)
\left[\partial_{\phi_i}\partial_{\phi_j}
\sum_{\pi \in \mathcal{P}(\mathcal{K})}\prod_{B\in\pi}\chi_{B}^{\phi}
\right]_{\boldsymbol{\phi} = \boldsymbol{0}}
\\
&= 
-\chi_l\chi_{ijk}
-\chi_k\chi_{ijl}
+ 
\left[\partial_{\phi_i}\partial_{\phi_j}
(\chi_{kl}^{\phi}+\chi_{k}^{\phi}\chi_{l}^{\phi})
\right]_{\boldsymbol{\phi} = \boldsymbol{0}}\\
&= 
-\chi_l\chi_{ijk}
-\chi_k\chi_{ijl}
+ 
\chi_{ijkl}
+\chi_{ik}\chi_{jl}
+\chi_{il}\chi_{jk}
+\chi_l\chi_{ijk}
+\chi_k\chi_{ijl}\\
&=\chi_{ijkl}+\chi_{ik}\chi_{jl}
+\chi_{il}\chi_{jk}\, .
\end{aligned}
\end{align*}

In summary, we recover the full $D(\boldsymbol{\chi})$ matrix, 
$$
D(\boldsymbol{\chi}) = 
\begin{pmatrix}
D_{i,k}
& D_{i,kl} \\
D_{ij,k}
& D_{ij,kl}
\end{pmatrix}
=
\begin{pmatrix}
\chi_{ik}
& \chi_{ikl} \\
\chi_{ijk}
& \chi_{ijkl}+\chi_{ik}\chi_{jl}+\chi_{il}\chi_{jk}
\end{pmatrix}\,,
$$
consistent with the expression in \eqref{eq:definition_D_2nd_QLE}\,.

\section*{Derivation of Explicit Expressions for First- and Second-Order cumulants under Pairwise Fitness Function}\phantomsection\label{sec:derivation_first_second_dynamis_SI}
\subsection*{Exact calculation}\phantomsection\label{sec:exact_calculation}
Here, we derive explicit expressions for the first- and second-order cumulants' equations of motion under the pairwise fitness function and demonstrate that the results derived from the exQLE yield the exact results.
Pairwise fitness is defined as:
\begin{align}
    \begin{aligned}
    F(g) &= \bar{F} + \sum_k s_k g_k + \sum_{k<l} s_{kl} g_k g_l\\
    &\bar{F} + \sum_k s_k g_k + \sum_{k<l} s_{kl} (g_k - \chi_k)(g_l - \chi_l) + \sum_{k<l} s_{kl} (g_k - \chi_k)\chi_l + \sum_{k<l} s_{kl} \chi_k (g_l - \chi_l) + \sum_{k<l} s_{kl} \chi_k \chi_l\,.
    \end{aligned}
\end{align}
Derivatives yield:
\begin{align}
\begin{aligned}
    \partial_{\chi_i} \langle F \rangle &= s_i + \sum_{l; l>i} s_{il} \chi_l + \sum_{k; k<i} s_{ki} \chi_k, \\
\partial_{\chi_{ij}} \langle F \rangle &= s_{ij}\,.
\end{aligned}
\end{align}
Subtracting,
\begin{equation}
F(g) - \langle F \rangle = \sum_k s_k (g_k - \chi_k) + \sum_{k<l} s_{kl} [(g_k - \chi_k)(g_l - \chi_l) + (g_k - \chi_k)\chi_l + \chi_k(g_l - \chi_l) - \chi_{kl}]
\end{equation}
Taking expectations:
\begin{align}
\begin{aligned}
    \langle F(g) - \langle F \rangle \rangle &= 0\,, \\
    \langle (g_i - \chi_i)(F(g) - \langle F \rangle) \rangle &= \sum_k s_k \chi_{ik} + \sum_{k<l} s_{kl} (\chi_{ikl} + \chi_{ik} \chi_l + \chi_k \chi_{il})\,, \\
    \langle (g_i - \chi_i)(g_j - \chi_j)(F(g) - \langle F \rangle) \rangle &= \sum_k s_k \chi_{ijk} + \sum_{k<l} s_{kl} (m_{ijkl} + \chi_{ijk} \chi_l + \chi_k \chi_{ijl} - \chi_{ij} \chi_{kl})\,.
\end{aligned}
\end{align}
where the fourth central moment is given by:
\begin{equation}
m_{ijkl} = \chi_{ijkl} + \chi_{ij} \chi_{kl} + \chi_{ik} \chi_{jl} + \chi_{il} \chi_{jk}.
\end{equation}
Therefore, the exact equations of motion are:
\begin{align}
\begin{aligned}
    \dot{\chi_i} &= \sum_k s_k \chi_{ik} + \sum_{k<l} s_{kl} (\chi_{ikl} + \chi_{ik} \chi_l + \chi_k \chi_{il})\,, \\
    \dot{\chi_{ij}} &= \sum_k s_k \chi_{ijk} + \sum_{k<l} s_{kl} (\chi_{ijkl} + \chi_{ik} \chi_{jl} + \chi_{il} \chi_{jk} + \chi_{ijk} \chi_l + \chi_k \chi_{ijl}) \,.
\end{aligned}
\end{align}

\subsection*{exQLE calculation}\phantomsection\label{sec:exQLE_calculation}
As the following calculations are valid for both the Gibbs distribution, and the cumulant distribution of any arbitrary distribution after taking the limit of $\boldsymbol{\phi}\to\boldsymbol{0}$, we assume that the genotype distribution takes the form of a Gibbs distribution for simplicity. 
For further simplicity, we drop $\boldsymbol{\phi}$ from $\chi_i^{\boldsymbol{\phi}},~ \chi_{ij}^{\boldsymbol{\phi}}$, and also omit the operation of taking the limit as $\boldsymbol{\phi\to 0}$\,.
The equation of motion for the first-order cumulant \eqref{eq:extended_qle_1st} is straightforward to obtain, using the relationships $\partial_{\phi_i}\chi_k=\chi_{ik} \text{ and } \partial_{\phi_i}\chi_{kl}=\chi_{ikl}$, which arise from the properties of the cumulant generating function or the normalization of the Gibbs distribution.  

Here, we focus on the equation of motion for the second-order cumulant. The direct calculation of the exQLE for the second-order cumulants yields \eqref{eq:extended_qle_2nd}, which is given as:  
\begin{align}
    \begin{aligned}
    \partial_{\phi_i}\partial_{\phi_j}\langle F \rangle 
    &=  \partial_{\phi_j}\left( 
    \sum_k(\partial_{\phi_i} \chi_k)\partial_{\chi_k}\langle F \rangle + 
    \sum_{k<l}(\partial_{\phi_i} \chi_{kl})\partial_{\chi_{kl}}\langle F \rangle
    \right)\\
    &=  
    \sum_k( \partial_{\phi_j}\partial_{\phi_i} \chi_k)\partial_{\chi_k}\langle F \rangle + 
    \sum_{k<l}( \partial_{\phi_j}\partial_{\phi_i} \chi_{kl})\partial_{\chi_{kl}}\langle F \rangle
    + 
    \sum_k(\partial_{\phi_i} \chi_k)\partial_{\phi_j}\partial_{\chi_k}\langle F \rangle + 
    \sum_{k<l}(\partial_{\phi_i} \chi_{kl})\cancelto{0}{\partial_{\phi_j}\partial_{\chi_{kl}}\langle F \rangle}\\
    &=  
    \sum_k \chi_{ijk}\ \partial_{\chi_k}\langle F \rangle + 
    \sum_{k<l}\ \chi_{ijkl}\partial_{\chi_{kl}}\langle F \rangle
    +
    \sum_k\chi_{ik} \left(\sum_l \chi_{jl}\partial_{\chi_l}\partial_{\chi_k}\langle F \rangle +
\sum_{l<m} \chi_{jlm}\cancelto{0}{\partial_{\chi_{lm}}\partial_{\chi_k}\langle F \rangle}\right)\\
    &=  
    \sum_k \chi_{ijk}\ \partial_{\chi_k}\langle F \rangle + 
    \sum_{k<l}\ (\chi_{ijkl}+\chi_{ik}\chi_{jl}+\chi_{il}\chi_{jk})\partial_{\chi_{kl}}\langle F \rangle\,.
    \end{aligned}
\end{align}
Here, we used the fact that the derivative of the average fitness beyond the second-order cumulant vanishes. Additionally we used:  $\partial_{\chi_k}\partial_{\chi_l}\langle F \rangle = \partial_{\chi_{kl}}\langle F \rangle$ and 
\begin{equation}
    \sum_{k,l}\chi_{ik}\chi_{jl}\partial_{\chi_{kl}}\langle F \rangle
    = \sum_{k<l}(\chi_{ik}\chi_{jl} + \chi_{il}\chi_{jk})\partial_{\chi_{kl}}\langle F \rangle\,.
\end{equation}
The equations of motion for exQLE, corresponding to \eqref{eq:extended_qle_1st} and \eqref{eq:extended_qle_2nd}, lead to:
\begin{align}
\begin{aligned}
    \dot{\chi_i} &= \sum_k \chi_{ik} \partial_{\chi_k} \langle F \rangle + \sum_{k<l} \chi_{ikl} \partial_{\chi_{kl}} \langle F \rangle \\
    &= \sum_k \chi_{ik} \left( s_k + \sum_{l;l>k} s_{kl} \chi_l + \sum_{l;l<k} s_{lk} \chi_l \right) + \sum_{k<l} s_{kl} \chi_{ikl} \\
    &= \sum_k \chi_{ik} s_k + \sum_{k<l} s_{kl} (\chi_{ik} \chi_l + \chi_{il} \chi_k + \chi_{ikl}) \,,
\end{aligned}
\end{align}
and 
\begin{align}
\begin{aligned}
    \dot{\chi_{ij}} &= \sum_k \chi_{ijk} \partial_{\chi_k} \langle F \rangle + \sum_{k<l} (\chi_{ijkl} + \chi_{ik} \chi_{jl} + \chi_{il} \chi_{jk}) \partial_{\chi_{kl}} \langle F \rangle \\
    &= \sum_k \chi_{ijk} \left( s_k + \sum_{l;l>k} s_{kl} \chi_l + \sum_{l;l<k} s_{lk} \chi_l \right) + \sum_{k<l} s_{kl} (\chi_{ijkl} + \chi_{ik} \chi_{jl} + \chi_{il} \chi_{jk}) \\
    &= \sum_k s_k \chi_{ijk} + \sum_{k<l} s_{kl} (\chi_{ijk} \chi_l + \chi_{ijl} \chi_k + \chi_{ijkl} + \chi_{ik} \chi_{jl} + \chi_{il} \chi_{jk})\, .
\end{aligned}
\end{align}
Here, where we used:
\begin{equation}
\sum_k \chi_{ik} \sum_{l;l<k} s_{lk} \chi_l = \sum_{k<l} s_{kl} \chi_{il} \chi_k\,,
\end{equation}
by simply renaming indices. The two sets of equations are identical.

Therefore, we directly confirmed that the equations of motion for the first and second cumulants, derived from the exact Price's equation and the exQLE equation, yield identical results in the case of pairwise fitness.

\section*{Proof of positive semidefiniteness of $D$ matrix for $K\in\{1,2\}$}
For the $K=1$ case, $D(\boldsymbol{\chi})$ is trivially positive semidefinite. 
For $K=2$ case, by defining $\Delta_i := g_i-\chi_i$, it can be expressed as 
\begin{align}
    \begin{aligned}
        D(\boldsymbol{\chi}) &= 
\begin{pmatrix}
\chi_{ik}
& \chi_{ikl} \\
\chi_{ijk}
& \chi_{ijkl}+\chi_{ik}\chi_{jl}+\chi_{il}\chi_{jk}
\end{pmatrix}\\
&=
\left\langle
\begin{pmatrix}
\Delta_i \Delta_j
& \Delta_i \Delta_k \Delta_l \\
\Delta_i \Delta_j \Delta_k
& (\Delta_i \Delta_j - \chi_{ij})
  (\Delta_k \Delta_l - \chi_{kl})
\end{pmatrix}
\right\rangle\\
&=
\left\langle
\begin{pmatrix}
\Delta_i \\
\Delta_i \Delta_j - \chi_{ij}
\end{pmatrix}
\begin{pmatrix}
\Delta_k \\
\Delta_k \Delta_l - \chi_{kl}
\end{pmatrix}^\top
\right\rangle \, \succeq 0\, .
    \end{aligned}
\end{align}
From the second line to the third line, we used the fact $\langle  \Delta_i \Delta_j \Delta_k \rangle 
= \langle  (\Delta_i \Delta_j - \chi_{ij}) \Delta_k  \rangle $\,. Therefore, $D(\boldsymbol{\chi})$ is positive semidefinite.

\section*{Derivation of the cumulant dynamics from genotype dynamics}\phantomsection\label{sec:genotype_dynamics_SI}
We denote the genotype distribution $P(\boldsymbol{g})$ for $\boldsymbol{g}\in \{-1,1\}^L$ as $P(\boldsymbol{g}^a) \mapsto z_a$, where each unique genotype is indexed by  $a\in\{1,\ldots, 2^L\}$\,.
We also denote genotype-level selection coefficient as $h_a$, such that the fitness function satisfies $F(\boldsymbol{g}^a) = \bar{F} + h_a$\,.

The average fitness is then given by
\begin{equation}
    \langle F \rangle 
= \bar{F} + \sum_a h_a z_a
= \bar{F} + \sum_{\mathcal{J}} s_{\mathcal{J}}\mu_{\mathcal{J}}
\,.
\end{equation}
Here, $\mu_\mathcal{J}$ denote a moment indexed by the set $\mathcal{J}$ (which may include multiple indices), and given by 
\begin{equation}
    \mu_{\mathcal{J}} = \sum_a z_a \prod_{j\in \mathcal{J}} g^a_j\,.
\end{equation}
Let define the matrix,
\begin{equation}
    \mathcal{G}_{a,\mathcal{J}} = \prod_{j\in \mathcal{J}} g^a_j\qquad,    
\end{equation}
so that the moments and genotype distributions are related via $\boldsymbol{\mu} = \mathcal{G}^\top \boldsymbol{z}$\,.

Given Fisher's fundamental theorem, the genotype distribution evolves as 
\begin{equation}
    \dot{z}_a = \sum_{ab} C_{ab}(\boldsymbol{z}) h_b\,
\end{equation}
where 
\begin{equation}
    C_{ab}(\boldsymbol{z}) = z_a\delta_{a,b} - z_a z_b\,,
\end{equation}
is the covariance matrix of genotype frequencies. 
In the vector form, it can be expressed as $\dot{\boldsymbol{z}} = C(\boldsymbol{z})\boldsymbol{h} = C(\boldsymbol{z})\boldsymbol{\nabla_z}\langle F \rangle$\,. 

Since $\langle F \rangle $ depends on $\boldsymbol{z}$ via the moments $\boldsymbol{\mu}$, we apply the chain rule:
\begin{equation}
    \frac{\partial \langle F \rangle}{\partial z_a} = \sum_{\mathcal{J}} \frac{\partial \mu_{\mathcal{J}}}{\partial z_a} \frac{\partial \langle F \rangle}{\partial \mu_{\mathcal{J}}} = \sum_{\mathcal{J}} \mathcal{G}_{a,\mathcal{J}} \frac{\partial \langle F \rangle}{\partial \mu_{\mathcal{J}}}\,,
\end{equation}
which can be expressed in vector form $\boldsymbol{\nabla_{z}}\langle F \rangle = \mathcal{G}\boldsymbol{\nabla_\mu}\langle F \rangle\,.$

Substituting into the dynamics:
\begin{equation}
    \dot{\boldsymbol{z}} = C(\boldsymbol{z})\mathcal{G}\boldsymbol{\nabla_{\mu}}\langle F \rangle \,.
\end{equation}
Using the relation $\boldsymbol{\mu} = \mathcal{G}^\top \boldsymbol{z}$, the moment dynamics become
\begin{equation}
    \dot{\boldsymbol{\mu}} = \mathcal{G}^\top C(\boldsymbol{z})\mathcal{G}\boldsymbol{\nabla_{\mu}}\langle F \rangle \,,
\end{equation}
which can be further transformed by using the chain rule:
\begin{equation}
    \dot{\boldsymbol{\mu}} = \mathcal{G}^\top C(\boldsymbol{z})\mathcal{G} \left( \boldsymbol{\nabla_\mu}\boldsymbol{\chi}^\top\right)\boldsymbol{\nabla_{\chi}}\langle F \rangle \,.
\end{equation}
Transforming from moments to cumulants using the Jacobian matrix: 
$\d\boldsymbol{\mu} = \left(\boldsymbol{\nabla_\chi}\boldsymbol{\mu}^\top\right)^\top \d \boldsymbol{\chi}$\,. 
Substituting this into the dynamics gives:
 $\dot{\boldsymbol{\mu}} = \left(\boldsymbol{\nabla_\chi}\boldsymbol{\mu}^\top\right)^\top \dot{\boldsymbol{\chi}}$. 
 
 By using the relation,
\begin{equation*}
    \left( \left(\boldsymbol{\nabla_\chi}\boldsymbol{\mu}^\top\right)^\top  \right)^{-1}
    =
    \left( \left(\boldsymbol{\nabla_\chi}\boldsymbol{\mu}^\top\right)^{-1}  \right)^\top
    =
    \left( \boldsymbol{\nabla_\mu}\boldsymbol{\chi}^\top  \right)^\top \,,
\end{equation*}
we obtain the cumulant dynamics:
\begin{equation}
    \dot{\boldsymbol{\chi}} = D(\boldsymbol{\chi}) \boldsymbol{\nabla_\chi}\langle F \rangle\,,
\end{equation}
where
\begin{equation}
    D(\boldsymbol{\chi}) = \left( \boldsymbol{\nabla_\mu}\boldsymbol{\chi}^\top\right)^\top\mathcal{G}^\top C(\boldsymbol{z})\mathcal{G} \left( \boldsymbol{\nabla_\mu}\boldsymbol{\chi}^\top\right)\, .
\end{equation}
This is the matrix $D(\boldsymbol{\chi})$ appearing in \eqref{eq:equation_generalized_QLE} of the main text. Its origin lies in $C(\boldsymbol{z})$, which arises from competition between genotypes and also serves as the covariance matrix in the diffusion process.
Importantly, since $C(\boldsymbol{z})$ is symmetric and positive semidefinite, the matrix $D(\boldsymbol{\chi})$ inherits these properties, it is symmetric and positive semidefinite as well.

\section*{Derivation of the equation to infer fitness values}\phantomsection\label{sec:derive_MLE_for_cumulants_SI}
We now derive the equation used to infer fitness parameters from cumulant dynamics.

The cumulant dynamics described in \eqref{eq:definition_D_2nd_QLE} have been deterministic, assuming an infinitely large population. For a finite population of size $N$, however, stochastic effects must be considered. In this case, the dynamics become a Langevin equation:
\begin{equation}\label{eq:LangevinSupp}
    \dot{\boldsymbol{\chi}} = D(\boldsymbol{\chi})\boldsymbol{\nabla_\chi} \langle F \rangle + \sqrt{D(\boldsymbol{\chi}(t))/N}\, \boldsymbol{\eta}(t)\,.
\end{equation}
where $\boldsymbol{\eta}(t)$ is a noise vector satisfying $\langle\boldsymbol{\eta}\rangle=\boldsymbol{0}$ and
$\langle \eta_{\mathcal{J}}(t) \eta_{\mathcal{K}}(t') \rangle = \delta_{\mathcal{J,K}}\delta(t-t')$\,.

This Langevin equation is equivalent to the following Fokker–Planck equation \cite{risken1989fokker}:
\begin{equation}
    \partial_t P(\boldsymbol{\chi},t) = - \boldsymbol{\nabla_\chi}^\top D(\boldsymbol{\chi}) \boldsymbol{\nabla_\chi}\langle F \rangle P(\boldsymbol{\chi}, t)+ {\small \frac{N}{2} }\Tr\left(\boldsymbol{\nabla_\chi}\boldsymbol{\nabla_\chi}^\top D(\boldsymbol{\chi}) \right) P(\boldsymbol{\chi}, t)\, .
\end{equation}

The Fokker–Planck equation can be rewritten as a probability density over entire cumulant trajectories. 
For numerical implementation, we discretize time at points $t_k$ for $k \in \{0, 1, \ldots, K+1\}$, and define
$\Delta t_k := t_{k+1} - t_k$ and $\boldsymbol{\Delta \chi}(t_k) := \boldsymbol{\chi}(t_{k+1}) - \boldsymbol{\chi}(t_k)$.

The probability of a cumulant trajectory is then expressed as
\begin{equation}
    P((\boldsymbol{\chi}(t_k))_{k=0}^{K+1} ) \propto e^{ - N\mathcal{S}((\boldsymbol{\chi}(t_k))_{k=0}^{K+1}) }\,, 
\end{equation}
where $\mathcal{S}$ is given by
\begin{align}
    \begin{aligned}
        \mathcal{S}((\boldsymbol{\chi}(t_k))_{k=0}^{K+1}) &= \sum_{k=0}^K \frac{1}{2\Delta t_k} \Big[ \boldsymbol{\Delta \chi}(t_k) - \Delta t_k D(\boldsymbol{\chi}(t_k)) \boldsymbol{\nabla_\chi}\langle F \rangle
    \Big]^\top \\
    &\qquad\times  D^{-1}(\boldsymbol{\chi}(t_k))
    \Big[ \boldsymbol{\Delta \chi}(t_k) - \Delta t_k D(\boldsymbol{\chi}(t_k))\boldsymbol{\nabla_\chi}\langle F \rangle
    \Big]\,.
    \end{aligned}
\end{align}
Since $\mathcal{S}$ is quadratic in ${\boldsymbol{\nabla_\chi}} \langle F \rangle$, the maximum likelihood estimate of the fitness parameters can be obtained analytically by solving the following linear equation:
\begin{equation}
    \sum_{k=0}^K {\Delta \boldsymbol{\chi}}(t_k)  = \sum_{k=0}^K 
    \Delta t_k D(\boldsymbol{\chi}(t_k)) 
    \boldsymbol{\nabla_\chi}\langle F \rangle \,.
\end{equation} 
By solving the above maximum likelihood equation, which is linear in $\boldsymbol{s}$, we can obtain $\boldsymbol{s}$\,.

\section*{Efficient computation of the forward exQLE simulation under higher-order fitness}\phantomsection\label{sec:efficient_computation_SI}
To obtain cumulant dynamics using either the exQLE framework or the exQLE with a Gaussian closure scheme, we must evaluate the products of the diffusion matrix, which involve third and fourth-order cumulants and fitness parameters.
However, a direct computation of the diffusion matrix and its product with fitness parameters is computationally expensive. Estimating these values across multiple time points, on the order of 100 different time points, requires significant computational time. Additionally, incorporating a fitness function that depends on four-way interactions further increases computational complexity.
To improve computational efficiency, we outline an optimized approach for computing cumulant dynamics.

Denote the gradient of the average fitness, which serves as the effective fitness parameter vector consisting of selection and epistatic coefficients, as  $\begin{pmatrix}
\hat{s}_k\\
\hat{s}_{kl}\\
\end{pmatrix}
:=
\begin{pmatrix}
\partial_{\chi_k}\\
\partial_{\chi_{kl}}
\end{pmatrix}\langle F \rangle$ \,.
For the fitness function with four-way interactions, let $\mu_i, \mu_{ij},\ldots $ denote mutation frequencies, and let $\langle F \rangle = \bar{F} + \sum_i s_i \mu_i + \sum_{i<j}s_{ij} \mu_{ij} + \sum_{i<j<k}s_{ijk} \mu_{ijk} + \sum_{i<j<k<l}s_{ijkl} \mu_{ijkl}$ represent the effective fitness parameters as
\begin{equation}\label{eq:effective_fitness_gradient_vector}
    \begin{pmatrix}
\hat{s}_k\\
\hat{s}_{kl}\\
\end{pmatrix}
    = \begin{pmatrix}
        s_k + \sum_{i(<k)}s_{ik}\mu_{i} + \sum_{i<j;~<k}s_{ijk}\mu_{ij} + \sum_{i<j<l~;<k}s_{ijlk} \mu_{ijl}\\
        s_{kl} + \sum_{i ;~<k<l}s_{ikl}\mu_{i} + \sum_{i<j;~<k<l}s_{ijkl}\mu_{ij}
    \end{pmatrix}\,.
\end{equation} These products between fitness parameters and moments can be computed as the sum of matrix-vector products, $s_{ijk}\mu_{ij}=\sum_{i<j(<k)}\langle s_{ijk}g_ig_j\rangle$ and $s_{ijlk} \mu_{ijl}=\sum_{i<j<l(<k)}\langle s_{ijlk}g_ig_jg_l\rangle$\,.

Let $\Delta_i := g_i - \chi_i$, then the cumulants can be obtained by $\chi_i = \langle \Delta_i \rangle\,,
\chi_{ij} = \langle \Delta_i \Delta_j \rangle\,,
\chi_{ijk} = \langle \Delta_i \Delta_j \Delta_k \rangle\,,
\chi_{ijkl} = \langle \Delta_i \Delta_j \Delta_k \Delta_l \rangle
- (\chi_{ik}\chi_{jl} + \chi_{il}\chi_{jk} + \chi_{ij}\chi_{kl})\,.$
Additionally, define $\chi^{(2)}$ as the second-order cumulant in matrix form and $\hat{S}$  as the effective epistasis matrix, where $\hat{s}_{kl}$ occupies the $k$-th row and $l$-th column.
Thus, the cumulant dynamics can be expressed in the following computationally more efficient form:
\begin{equation}\label{eq:cumulant_dynamics_efficient}
    \begin{pmatrix}
\dot{\chi}_i\\
\dot{\chi}_{ij}
\end{pmatrix}
= 
\begin{pmatrix}
\chi_{ik}
& \chi_{ikl} \\
\chi_{ijk}
& \chi_{ijkl}+\chi_{ik}\chi_{jl}+\chi_{il}\chi_{jk}
\end{pmatrix}
\begin{pmatrix}
\hat{s}_k\\
\hat{s}_{kl}\\
\end{pmatrix}
= 
\begin{pmatrix}
\left\langle \boldsymbol{\delta} (\boldsymbol{\Delta}^\top \boldsymbol{\hat{s}})\right\rangle
+ \left\langle \boldsymbol{\Delta} (\boldsymbol{\Delta}^\top \hat{S} \boldsymbol{\Delta})\right\rangle/2
\\
\left \langle \boldsymbol{\Delta}\boldsymbol{\Delta}^\top  (\boldsymbol{\Delta}^\top\boldsymbol{\hat{s}})\right\rangle
+ 
\left \langle \boldsymbol{\Delta} \boldsymbol{\Delta}^\top \boldsymbol{\Delta}^\top \hat{S} \boldsymbol{\Delta}\right\rangle - \chi^{(2)}\mathrm{Sum}\left(\chi^{(2)} \odot \hat{S}\right)
\end{pmatrix}\,.
\end{equation}
The last expression in \eqref{eq:cumulant_dynamics_efficient} is efficient because this computation never explicitly requires obtaining matrices or matrix products with more than $\mathcal{O}(L^2)$ elements. 

For the Gaussian closure scheme, the cumulant dynamics can be obtained in the same manner. The dynamics under the Gaussian closure scheme are given by:
\begin{align}
    \begin{aligned}
        \begin{pmatrix}
\dot{\chi}_i\\
\dot{\chi}_{ij}
\end{pmatrix}
&= 
\begin{pmatrix}
\chi_{ik}
& \chi_{ikl}(\delta_{ik}+\delta_{jk}) \\
\chi_{ijk}(\delta_{ik}+\delta_{jk})
& \chi_{ijkl}\delta_{ik}\delta_{jl}+\chi_{ik}\chi_{jl}+\chi_{il}\chi_{jk}
\end{pmatrix}
\begin{pmatrix}
\hat{s}_k\\
\hat{s}_{kl}\\
\end{pmatrix}\\
&= 
\begin{pmatrix}
\chi_{ik}
& \chi_{ikl}(\delta_{ik}+\delta_{jk}) \\
\chi_{ijk}(\delta_{ik}+\delta_{jk})
& [\langle \Delta_i \Delta_j \Delta_i \Delta_j\rangle 
-\chi_{ii}\chi_{jj}-2\chi_{ij}^2
]\delta_{ik}\delta_{jl}+\chi_{ik}\chi_{jl}+\chi_{il}\chi_{jk}
\end{pmatrix}
\begin{pmatrix}
\hat{s}_k\\
\hat{s}_{kl}\\
\end{pmatrix}\\
&= 
\begin{pmatrix}
\langle \boldsymbol{\Delta} (\boldsymbol{\Delta}^\top \boldsymbol{\hat{s}})\rangle + 
\left\langle  \boldsymbol{\Delta} \odot
\left(
\sum_{k}\left\{\left(\boldsymbol{\Delta}\boldsymbol{\Delta}^\top \odot \hat{S}\right)_{ik} +
\left(\boldsymbol{\Delta}\boldsymbol{\Delta}^\top \odot \hat{S}\right)_{ki}
\right\}
\right)_{i=1}^L
\right\rangle
 \\
\left \langle \boldsymbol{\Delta} \boldsymbol{\Delta} ^\top \odot \left(\boldsymbol{1}(\boldsymbol{\Delta} \odot \boldsymbol{\hat{s}})^\top + (\boldsymbol{\Delta} \odot \boldsymbol{\hat{s}})\boldsymbol{1}^\top  \right) \right\rangle
+ \mathrm{diag}\left(
\left(
\left(
\langle \Delta_i^2\Delta_j^2 \rangle
-\chi_{ii}\chi_{jj}-2\chi_{ij}^2
\right)s_{ij}
\right)_{i<j}
\right) + \chi^{(2)} \hat{S} \chi^{(2)} 
\end{pmatrix}\,, 
    \end{aligned}
\end{align}
where $\boldsymbol{1}$ is a vector of length $L$ consisting entirely of ones, $\odot$ denotes the elementwise product, and the notation $(a_i)_i^L=(a_1, \ldots, a_L)^\top$ is used.

\clearpage
\begin{figure*}
\centering
\includegraphics[width=0.5\linewidth]{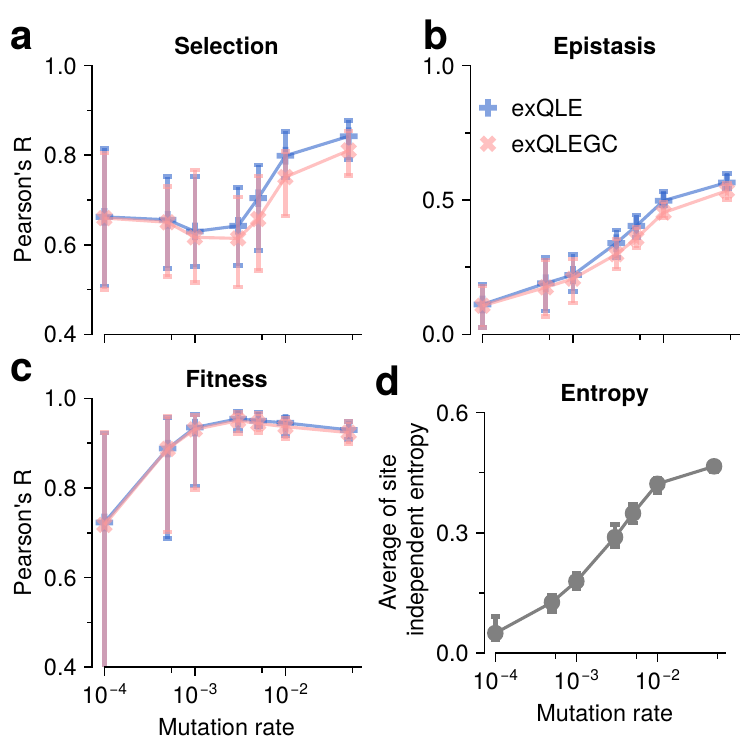}
\caption{ \label{fig:SI_Pearson_HI}
\textbf{Accuracy of inferred fitness parameters under higher-order fitness function.}
This corresponds to \textbf{Fig.~\ref{fig:accuracy_PW}} in the main text, but here the underlying selective pressure is determined by a higher-order fitness function rather than a pairwise one, that is, $K^* = 4$, where $K^*$ denotes the highest order of cumulants in the averaged fitness function (as defined in the main text). The functionality and model parameters are the same as those used in the simulation for \textbf{Fig.~\ref{fig:rates_of_changes_in_1st_and_2nd_order_cumulant}}.
The inference approach remains pairwise ($K = 2$), aiming to infer additive (selection) and pairwise (epistatic) fitness parameters. Despite the increased complexity of the true fitness landscape, the inferred parameters show high accuracy, as measured by Pearson’s $R$ values, for selection, epistasis, and overall fitness. The dependency of accuracy on mutation rate remains qualitatively similar to that observed under the pairwise fitness setting.
\label{fig:Pearson_HI}
}
\end{figure*}

\newpage
\begin{figure*}
\centering
\includegraphics[width=0.5\linewidth]{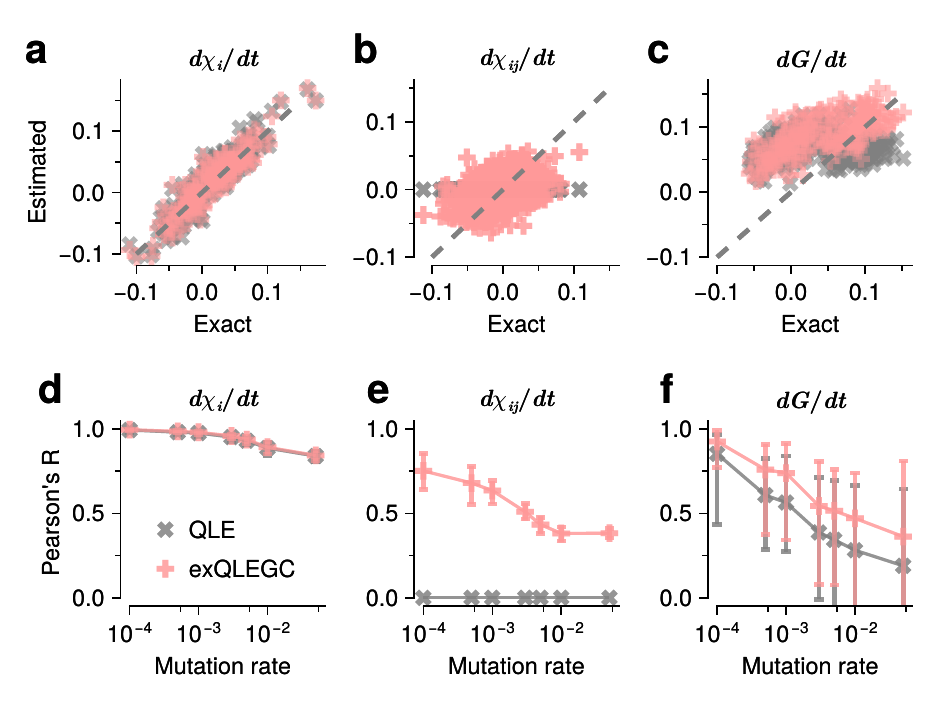}
\caption
{ \label{fig:SI_dynamics_exQLEGC}
\textbf{ Dynamics of cumulants and traits using exQLE and Gaussian closure scheme.}
This corresponds to \textbf{Fig.~\ref{fig:rates_of_changes_in_1st_and_2nd_order_cumulant}} in the main text but utilizes the exQLE framework with a Gaussian closure (GC) scheme, which suppresses all cumulants beyond second order. The GC scheme is efficient for inferring fitness parameters and is not limited to inference problems; it is also applicable to forward processes. Although the accuracy of the dynamics for cumulants (\textbf{b}) is not as high as that in the exQLE case without GC, the estimated values show a reasonable correlation with those from the exact calculations. 
Pearson's R values for additive selection (\textbf{a}), pairwise epistasis (\textbf{b}), and random traits (\textbf{c}) for exQLE with GC  are $0.93, 0.43$, and $0.65$. For comparison, the exQLE values without GC are $0.97, 0.89$, and $0.95$, respectively.
\label{fig:rate_of_changes_QLEGC}
}
\end{figure*}

\clearpage
\newpage
\section*{References}
\bibliographystyle{apsrev4-2}
\bibliography{reference.bib}